# A Multi-Tiered Bayesian Network Coastal Compound Flood Analysis Framework

**Preprinted manuscript. Under review.**

Ziyue Liu, Ph.D.,[1] Meredith L. Carr, Ph.D., P.E.,[2] Norberto C. Nadal-Caraballo, Ph.D.,[2] Luke A. Aucoin,[2] Madison C. Yawn,[2] and Michelle T. Bensi, Ph.D.[3]

[1]Industrial Engineering & Political Science, Purdue University; email: liu4205@purdue.edu; ORCID: 0009-0003-5850-0374
[2]Coastal & Hydraulics Laboratory, U.S. Army Engineer R&D Center;
[3]Department of Civil and Environmental Engineering, University of Maryland;

## Abstract

Coastal compound floods (CCFs) are triggered by the interaction of multiple mechanisms, such as storm surges, storm rainfall, tides, and river flow. These events can bring significant damage to communities, and there is an increasing demand for accurate and efficient probabilistic analyses of CCFs to support risk assessments and decision-making. In this study, a multi-tiered Bayesian network (BN) CCF analysis framework is established. In this framework, conceptual designs of multiple tiers of BN models with varying complexities are developed for application with varying levels of data availability and resources. A case study is conducted in New Orleans, LA, with three tiers of BN models constructed to demonstrate this framework. In the Tier-1 BN model, storm surges and river flow are incorporated based on hydrodynamic simulations. A seasonality node is used to capture the dependence between concurrent river flow and tropical cyclone (TC) parameters. In the Tier-2 BN model, joint distribution models of TC parameters are built for separate TC intensity categories. TC-induced rainfall is modeled as input to hydraulic simulations. In the Tier-3 BN model, potential variations of meteorological conditions are incorporated by quantifying their effects on TC activity and coastal water level. Flood antecedent conditions are also incorporated to more completely represent the conditions contributing to flood severity. In this case study, a series of joint distribution, numerical, machine learning, and experimental models are used to compute conditional probability tables needed for the BNs. A series of probabilistic analyses is performed based on these BN models, including CCF hazard curve construction and CCF deaggregation. The results of the analysis demonstrate the promise of this framework in performing CCF hazard analysis under varying levels of resource availability and project needs.

## 1. Introduction

Coastal regions are vulnerable to coastal compound floods (CCFs) that result from the combined mechanisms of storm surges, tides, intense rainfall, and (in some cases) river flooding. CCFs can bring substantial losses to communities, and, as a result, there is an increasing demand for accurate and efficient probabilistic analyses of CCFs to support risk assessments and decision-making processes.

Xu et al. (2022) documented a literature review of the development of CCF studies over the past two decades. The literature review covered CCF mechanism analysis, research methodology, and probabilistic analysis. They found that there is limited literature integrating statistical and numerical models for CCF hazard analysis. Zhang and Najafi (2020) provided a literature review of probabilistic CCF analysis in conjunction with a probabilistic case study on the southern coast of Saint Lucia in the Caribbean Sea. Their work demonstrates the potential of using simplified hydrodynamic models for CCF simulations. Bensi et al. (2020) provided a review of a range of studies addressing compound flooding hazards and identified several main classes of studies, including those that seek to characterize the joint distribution of multiple severity measures associated with CCFs (e.g., using parametric, non-parametric, and copula-based joint distributions) and to apply Bayesian approaches.

In the past decade, the US Army Corps of Engineers (USACE) has undertaken a series of efforts to probabilistically analyze hazards along coastal regions in the US and has developed the Coastal Hazard System Probabilistic Framework (CHS-PF) with Joint Probability Method Augmented by Metamodel Prediction (JPM-AMP) (Melby et al. 2017; Nadal-Caraballo et al. 2015, 2020, 2022). In the CHS-PF, a joint probability method integral equation is used to compute the frequency of exceedance of tropical cyclone (TC) induced coastal hazards:

$$\lambda_{R>r} = \lambda \int P(R > r|\mathbf{x}, \varepsilon)\, f_{\mathbf{X}}(\mathbf{x}) f_{\varepsilon}(\varepsilon) d\mathbf{x} \mathrm{d}\varepsilon \tag{1}$$

where $\lambda_{R>r}$ is the annual exceedance frequency (AEF) of storm response ($R$) exceeding a threshold ($r$) due to TC events. $\lambda$ is the storm recurrence rate; $P(R > r|\mathbf{x}, \varepsilon)$ is the conditional probability of $R > r$ given TC parameter forcing vector $\mathbf{x}$ and error $\varepsilon$. $f_{\mathbf{X}}(\mathbf{x})$ is the probability density function (PDF) of TC parameter forcing vector $\mathbf{x}$. $f_{\varepsilon}(\varepsilon)$ is the PDF of error $\varepsilon$.

This study explores the use of Bayesian networks (BNs) to enable the extension of the CHS-PF to support hazard analysis of CCFs. A BN is a Bayesian rule-motivated graphical model that represents the probabilistic relationships among multiple dependent variables. It excels in visualizing the dependence among variables and supporting probabilistic inference involving complex dependencies.

To illustrate key concepts associated with BNs, Figure 1 presents a BN with four fully dependent variables. In a BN, each node represents a variable, and each arrow (link) represents the (typically causal) relationship from a “parent” node to a “child” node. For example, in the BN shown in Figure 1, $X_1$ is a parent node of $X_2$, $X_3$ and $X_4$, while $X_4$ is a child node of $X_1$, $X_2$ and $X_3$. To build a BN model that can be implemented for probabilistic analysis, every variable typically needs to be discretized (with the exception of selected special cases). A conditional probability table (CPT) is assigned to each node. A CPT provides the conditional probability that a node $X_i$ takes on each of its possible states given each possible combination of states of its parent nodes. Additional information regarding BNs can be found in a range of references (e.g., Kjaerulff and Madsen 2008).

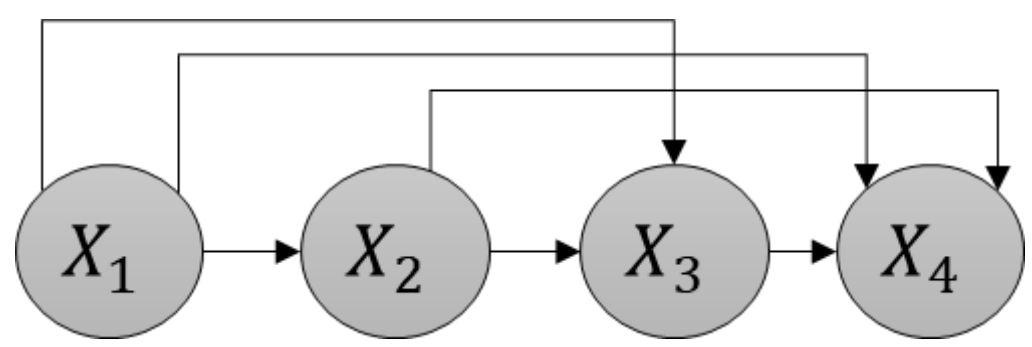

**Figure 1: Example BN with four fully dependent variables**

This study conceptualizes and constructs three tiers of BN models. These three tiers of BN models have different complexities and are designed for cases with varying levels of data and computational resource availability and project needs. The BN-based CCF modeling approach builds upon the proof-of-concept BN presented in Mohammadi et al. (2023) and subsequent BN-based applications in Liu et al. (2025). Mohammadi et al. (2023) built a BN for a coastal compound flood case study for a site located on the Delaware River near Trenton, New Jersey, where storm surges, storm precipitation, river flow, and tides have been included. In Liu et al. (2025), a BN built based on USACE’s CHS-PF has been implemented to develop a deaggregation method for multiple coastal hazards without addressing the interaction of coastal hazards.

Figure 2 presents conceptual designs of the three tiers of BN models proposed in this study. In the Tier-1 BN (Figure 2 (a)), which reflects an approach similar to that used in Nadal-Caraballo et al. (2022), storm surges and river flow are treated as the dominant CCF drivers. The dependence between storm events and river flow is modeled through seasonality during hurricane seasons. In the Tier-2 BN (Figure 2 (b)), storm rainfall is added as a contributor to CCFs. In the Tier-3 BN (Figure 2 (c)), variations in meteorological conditions and antecedent river conditions are added to build a more complex CCF model with more comprehensive mechanisms represented in coastal compound hazards. It is important to note that although the three tiers of the BN are designed to probabilistically analyze CCF with increasing complexity, the accuracy of each tier is also influenced by the selected statistical and physical models used to construct the respective BN.

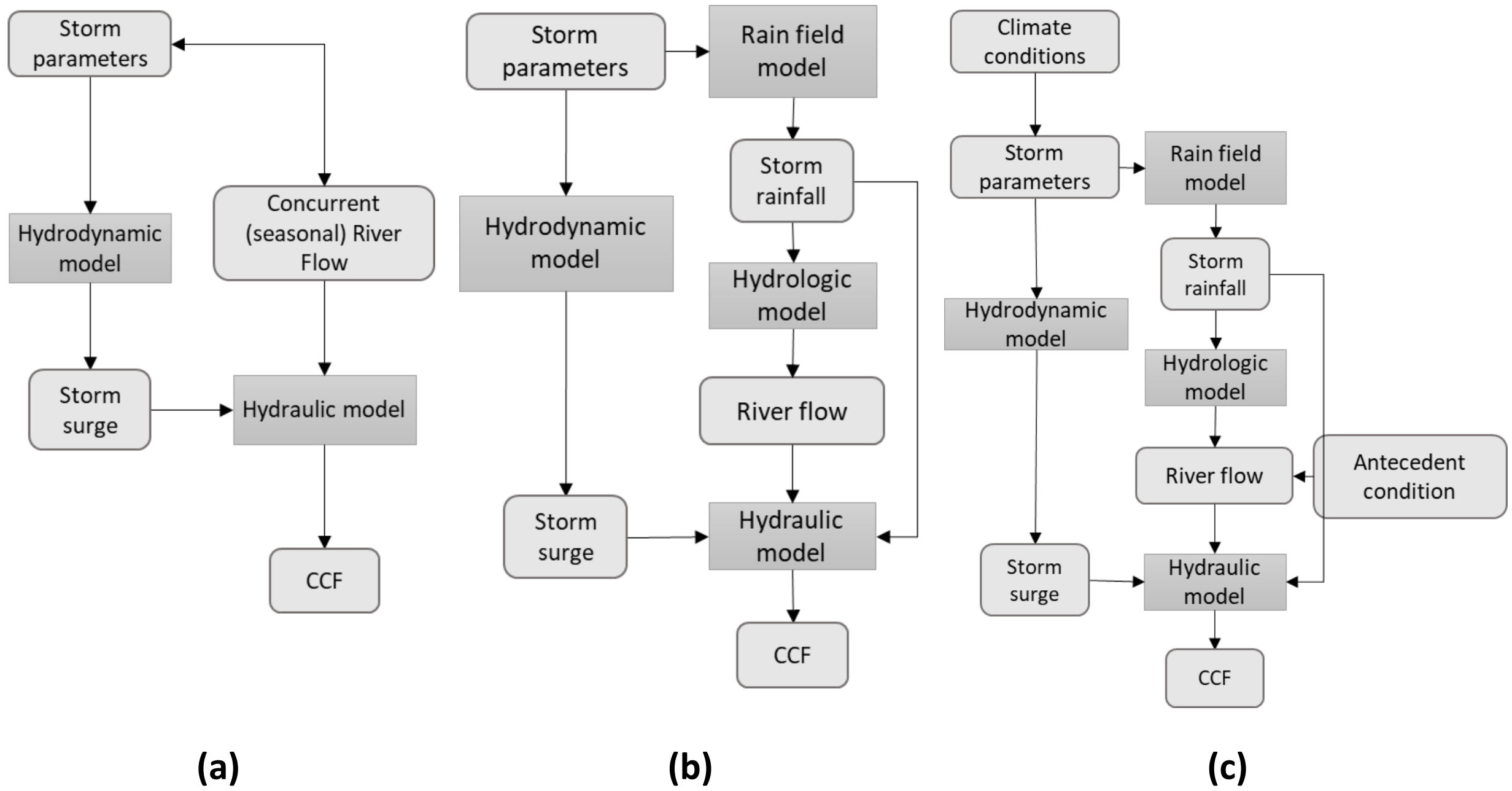

**Figure 2: Conceptual designs of three tiers of BN models: (a) Tier-1 BN; (b) Tier-2 BN; (c) Tier-3 BN**

To illustrate the proposed framework, these three tiers of BN are implemented using a case study in the region of New Orleans, LA. Two study sites (shown in Figure 3) are selected in the transition zone, where both surge and river flow impact water levels. The two study sites are selected also considering that they are influenced by hydraulic structures and can be reliably simulated with the employed models and assumptions. The Mississippi River (MR) is treated as the river flow contributor to CCF in the study region. While the case study is used to illustrate the framework for a representative set of assumptions and models, the multi-tiered BN approach itself is intended to be valid more generally.

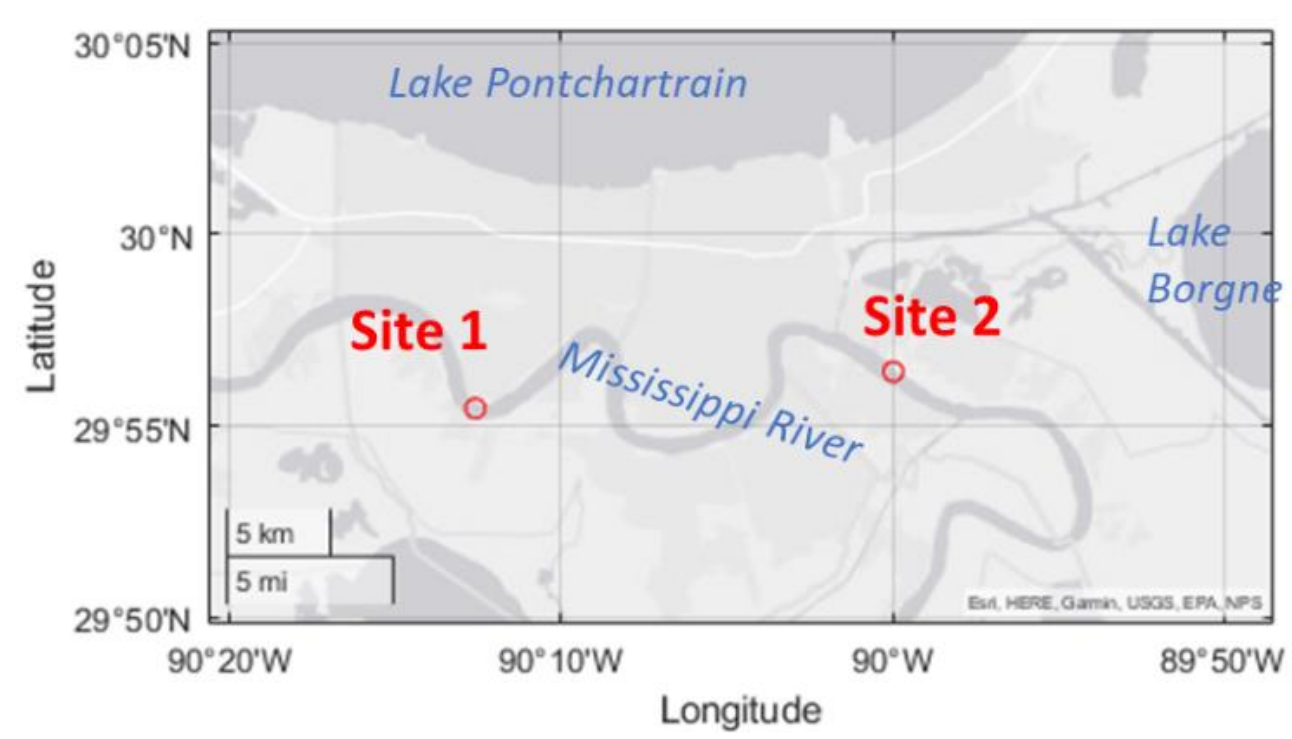

**Figure 3: Locations of study sites**

## 2. Tier-1 Bayesian Network Construction

A Tier-1 BN model is constructed leveraging the modeling approach first employed in the USACE's CHS-LA (Coastal Hazards System–Louisiana) study (Nadal-Caraballo et al. 2022). Hydrological seasonality is incorporated in this BN to capture the dependence between TC atmospheric parameters and concurrent river flow. Typically, seasonal events like snowmelt and rainfall will increase river discharge. Meanwhile, varying temperatures and moisture can affect the formation of storms (Gray 1968; Palmen 1948). Michele and Salvadori (2002) use river seasonality to capture flood antecedent conditions. Walega and Młyński (2017) evaluated the seasonality discharge of selected Carpathian rivers and found that the interaction of spring thaw and summer rain can significantly impact local river flow.

Figure 4 presents this Tier-1 BN, and Table 1 lists the physical definition of each node. The discretization information of each node is provided in Appendix A. In this study, NAVD88 2009.65 is used as the datum of all water surface levels (WSL). This Tier-1 BN consists of three primary models:

- **Seasonality-dependent river flow model** (as identified by black links in Figure 4): A seasonality node $S$, with its states representing the time period within the hurricane season, is used to capture the statistical relationship between TC atmospheric parameters (nodes $\Delta P$, $V_\mathrm{f}$, $R_\mathrm{max}$, and $\Theta$) and concurrent river flow (node $Q$). As reflected by the fact that node $S$ serves as a common parent of nodes $\Delta P$, $V_\mathrm{f}$, $R_\mathrm{max}$, $\Theta$, and $Q$.
- **Seasonality-dependent joint probability model for TC atmospheric parameters** (as identified by blue links in Figure 4): Joint probability models for TC atmospheric parameters are constructed for each season separately, and the dependencies among TC atmospheric parameters is reflected by the series of links among them.
- **Hydrodynamic-model-derived water surface level model** (as identified by red links in Figure 4): The estimated CCF WSL resulting from storm surges and river flow calculated in ADCIRC is represented by node $\hat{Z}$, which is a child of nodes $Q$, $\Delta P$, $V_\mathrm{f}$, $R_\mathrm{max}$, $\Theta$ and $X_0$. The associated actual CCF WSL is a function of the estimated value and an error term, as reflected by node $Z$ with parent nodes $\hat{Z}$ and $\varepsilon_Z$. Finally, the effects of astronomical tides are included through the addition of a tidal residual (represented by node $T$), resulting in a final WSL, as reflected by node $Z_\mathrm{t}$ with parent nodes $Z$ and $T$.

Additional details regarding the generation of CPTs associated with each of the above models are provided in the subsections that follow.

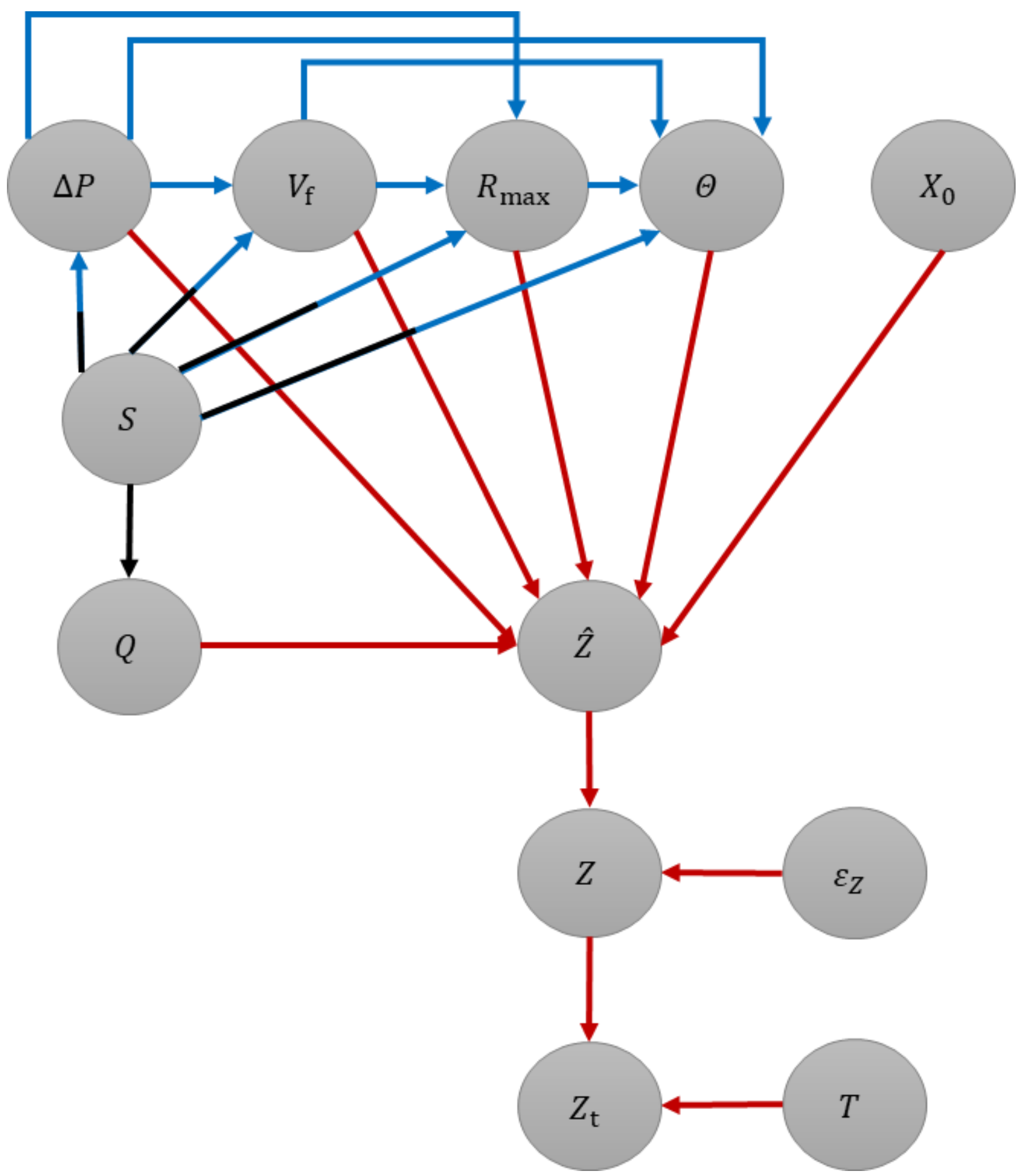

**Figure 4: Tier-1 BN (half-black, half-blue links represent relationships captured by two models)**

**Table 1. Physical definition of each node in Tier-1 BN**

| Notation | Physical definition | Notation | Physical definition |
|---|---|---|---|
| $S$: | Time period within hurricane season | $Q$: | MR river flow (kcfs) |
| $\Delta P$: | Storm central pressure deficit (hPa) | $\hat{Z}$: | WSL (m) resulting from storm surges and river flow predicted by model |
| $V_f$: | Storm forward velocity (km/hr) | $Z$: | WSL (m) resulting from storm surges and river flow with model error considered |
| $R_{\max}$: | Storm radius to maximum winds (km) | $\varepsilon_Z$: | Normalized residuals of WSL error |
| $\Theta$: | Storm heading direction (deg) | $T$: | Normalized residuals of astronomical tides |
| $X_0$: | Storm landfall location index | $Z_{\mathrm{t}}$: | WSL (m) with astronomical tides considered |

## 2.1. Seasonality-dependent river flow model

To model the dependence between storm parameters and concurrent (seasonal) river flow (as described in Figure 2 (a)), a seasonality node is set as a parent node for both river flow nodes ($Q$) and TC atmospheric parameters nodes ($\Delta P$, $V_{\mathrm{f}}$, $R_{\max}$, and $\Theta$). The seasonality node $S$ is discretized into time periods within the hurricane season, and different distributions for river flow and TC atmospheric parameters are defined for each season. In this case study, two time periods are used: Season 1 (May 1 to August 31) and Season 2 (September 1 to November 30). The dividing point between "meteorological summer" and "meteorological fall" (i.e., September 1) is subjectively selected for this case study. A weighting strategy is used to mitigate the effect of the hard breakpoint that occurs when dividing time series data. For each season, a probability is assigned based on the portion of observed storms that initiate inside that season. In this study, this resulted in $p(\text{Season 1}) = 0.46$, and $p(\text{Season 2}) = 0.54$.

The daily discharge data of United States Geological Survey (USGS) gage 07295100 at Red River Landing (Tarbert Landing) is used to compute the CPT of the MR river flow node $Q$. The gauge location is shown in Figure 5. Consistent with Nadal-Caraballo et al. (2022), it is assumed that there is a 70% and 30% split in discharge directed through the Mississippi and Atchafalaya Rivers, respectively, downstream of USGS gage 07295100. Gauge records from 1976 to 2019 and inside the window of hurricane season (May 1-November 30) are used. The MR daily discharge is regulated to a maximum of 35396 $m^3/s$ (1250 kcfs). The temporal record of MR daily discharge data is shown in Figure 6, with the red line indicating the subjectively selected meteorological breakpoint between seasons.

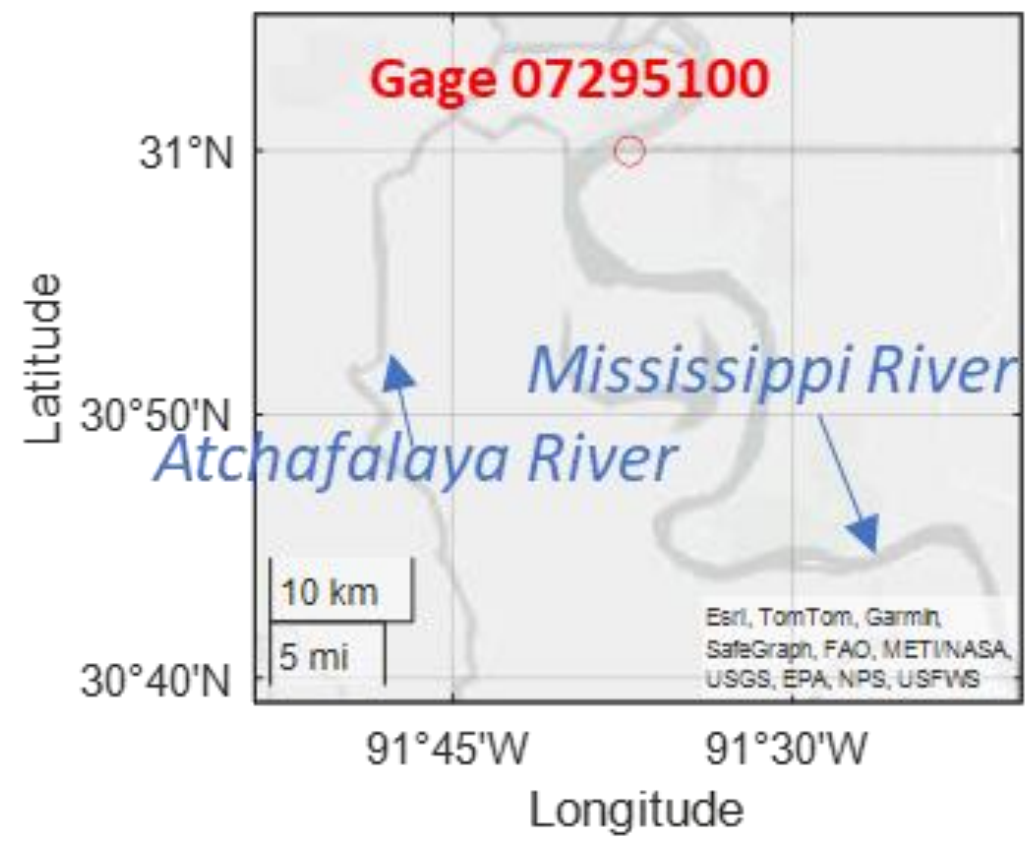

**Figure 5: Location of USGS gage 07295100 (about 320 km upstream of Site 1)**

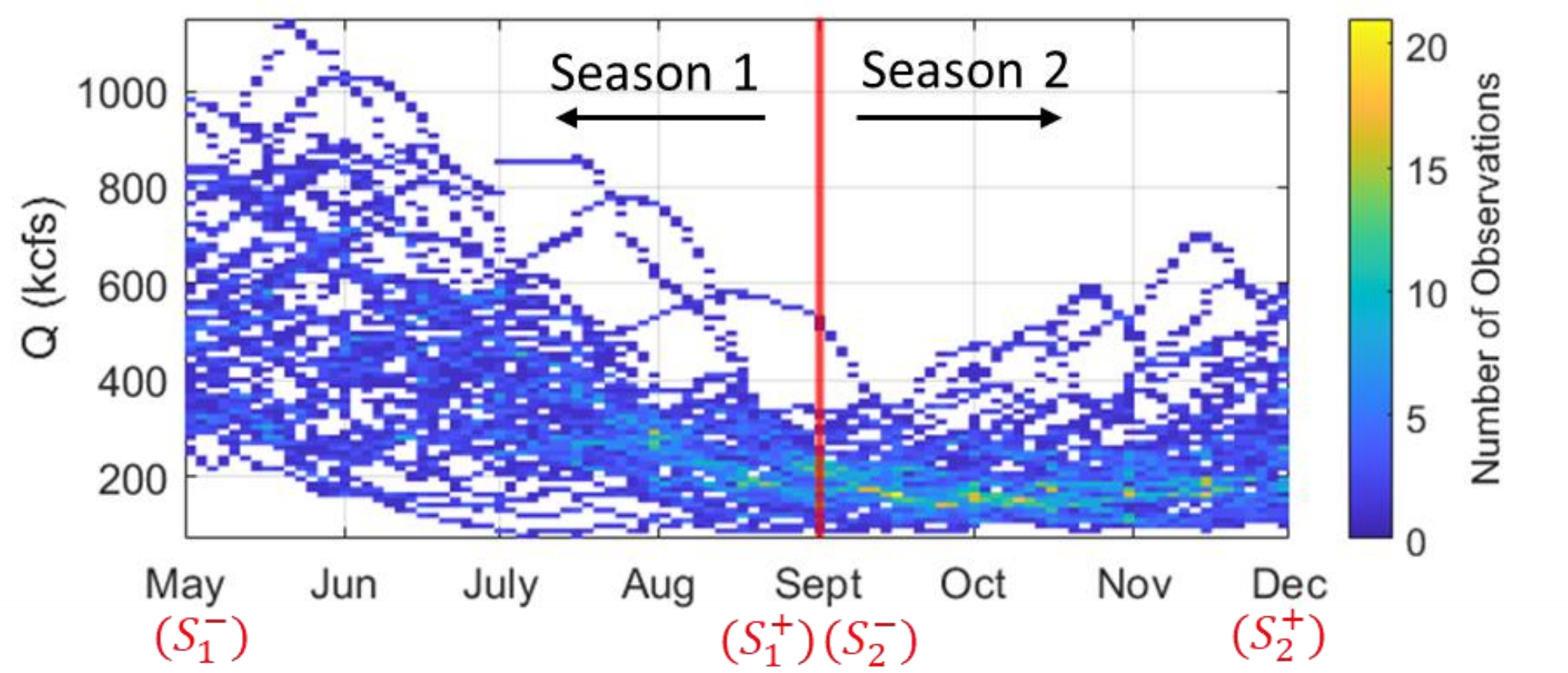

**Figure 6: Temporal distribution of sample data of MR daily discharge with a red line depicting the dividing point between Season 1 and Season 2**

To develop the conditional distributions of river flow for each season, historical daily river flow data are processed into independent statistical samples for each season period. First, to mitigate the effect of the autocorrelation in the time series data, river daily discharge data within hurricane season (i.e., records within the window of May 1 to November 30) are randomly selected. Considering that river flow data exhibits gradual seasonal transitions, a Gaussian kernel function (GKF) based temporal distance weighting method is used here to smooth out the effect of the season division in the time series. To generate the statistical samples for a season, each selected record is assigned a temporal distance $d_i$ to the boundaries of the corresponding Season $k$. The temporal distance is calculated as:

$$d_i = \begin{cases} S_k^- - t_i, & \text{if } t_i < S_k^- \\ 0, & \text{if } S_k^- \leq t_i \leq S_k^+ \\ t_i - S_k^+, & \text{if } S_k^+ \leq t_i \end{cases} \tag{2}$$

where $S_k^-$ is the Julian date of the first day of the season $k$, $S_k^+$ is the Julian date of the last day of Season $k$, $t_i$ is the date of the $i$th record.

For each record, a temporal distance weight is calculated using a GKF:

$$w_t(d_i) = \frac{1}{\sqrt{2\pi}h_d} \exp\left[-\frac{1}{2}\left(\frac{d_i}{h_d}\right)^2\right] \tag{3}$$

where $w_t(d_i)$ is the temporal distance weight of the $i$th record and $h_d$ is the kernel bandwidth which set as $h_d = 15$ days in this study. Using this method, an identical highest weight is assigned for each data point within the seasonal time window, and the assigned weight decreases as the distance between the data and the time window increases, as illustrated in Figure 7.

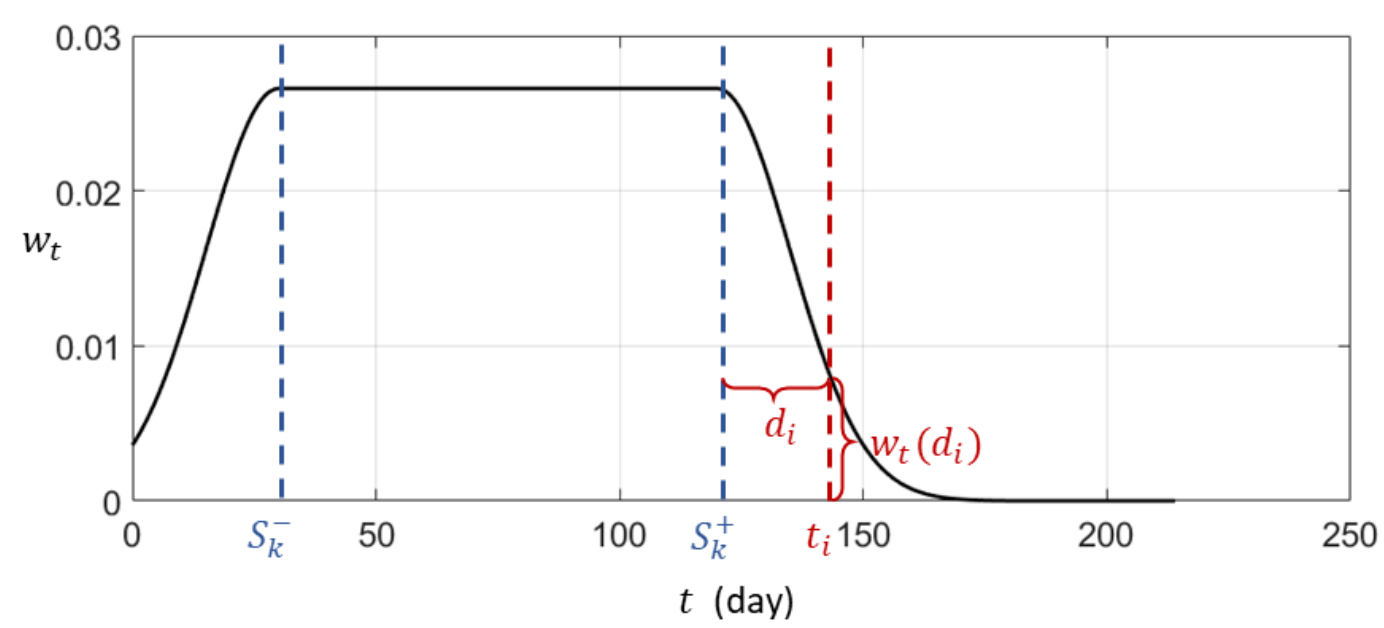

**Figure 7: Example of assigned weight along the time axis**

After calculating the temporal distance weight using the method described above, a distance weight adjustment method (Chouinard and Liu 1997; Nadal-Caraballo et al. 2015) is applied to selected records to obtain river flow statistical samples. Given the representative assumptions and models employed in the case study, Figure 8 plots the fitted marginal PDF curves with the corresponding samples of MR daily discharge in each season. Generalized extreme value distributions (MathWorks 2022) are used to fit marginal distributions of MR daily discharge in each season.

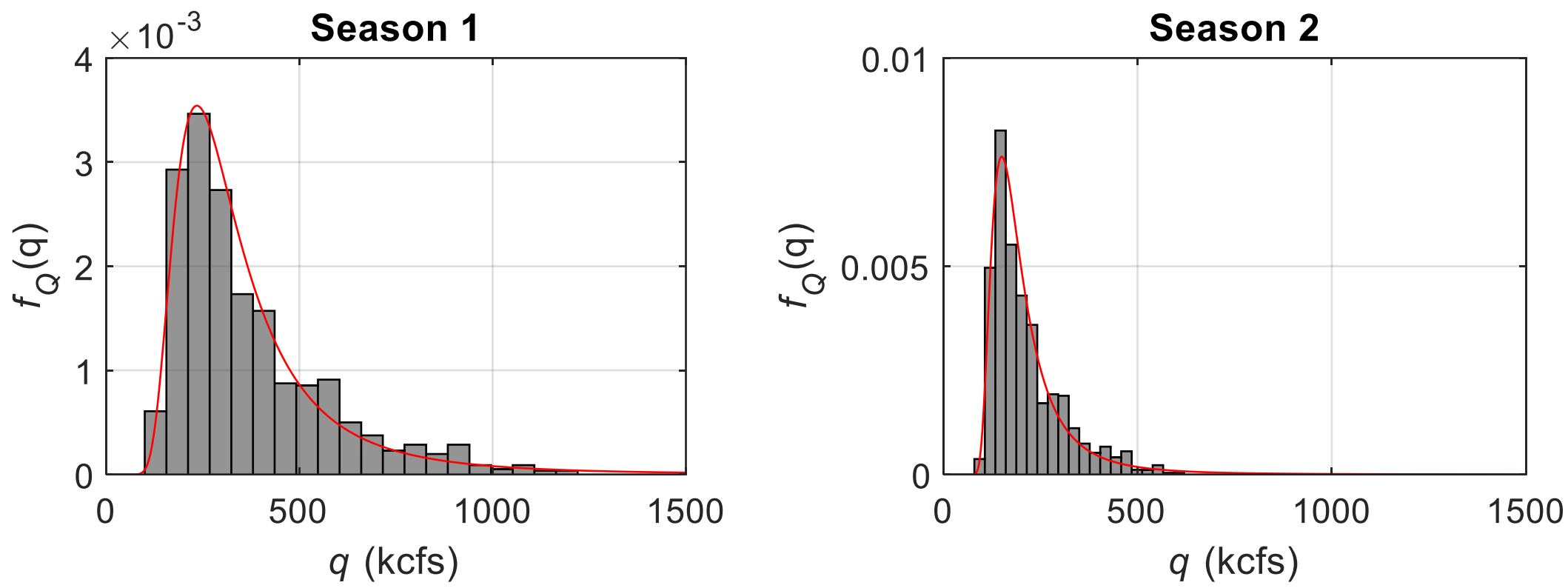

**Figure 8: Fitted marginal PDF curves with the corresponding statistical samples of MR daily discharge; Note scales of y-axis are different in each plot**

## 2.2. Seasonality-dependent joint probability model for TC atmospheric parameters

The Seasonality node $S$ is also connected to TC atmospheric parameters nodes ($\Delta P$, $V_{\mathrm{f}}$, $R_{\max}$, and $\Theta$), and separate joint probability models are built for each season. Those joint probability models are generated based on a statistical analysis of historical data. A historical storm dataset combining Hurricane Database 2 (HURDAT2) (Landsea and Franklin 2013; NOAA 2022) and Tropical Cyclone Extended Best Track Dataset (EBTRK) database (Demuth et al. 2006; RAMMB/CIRA 2021) with data imputation (Liu et al. 2024c) is used to generate storm statistical sampling data following the storm statistical sampling procedure of CHS-PF (Chouinard and Liu 1997; Nadal-Caraballo et al. 2015). A coastal reference location (CRL) with latitude = 29.576 and longitude = -89.538 and a 600 km radius capture zone is employed for this procedure (see Figure 9). The storm statistical sampling data are partitioned for each season and used to perform a seasonally-dependent joint distribution analysis using copulas.

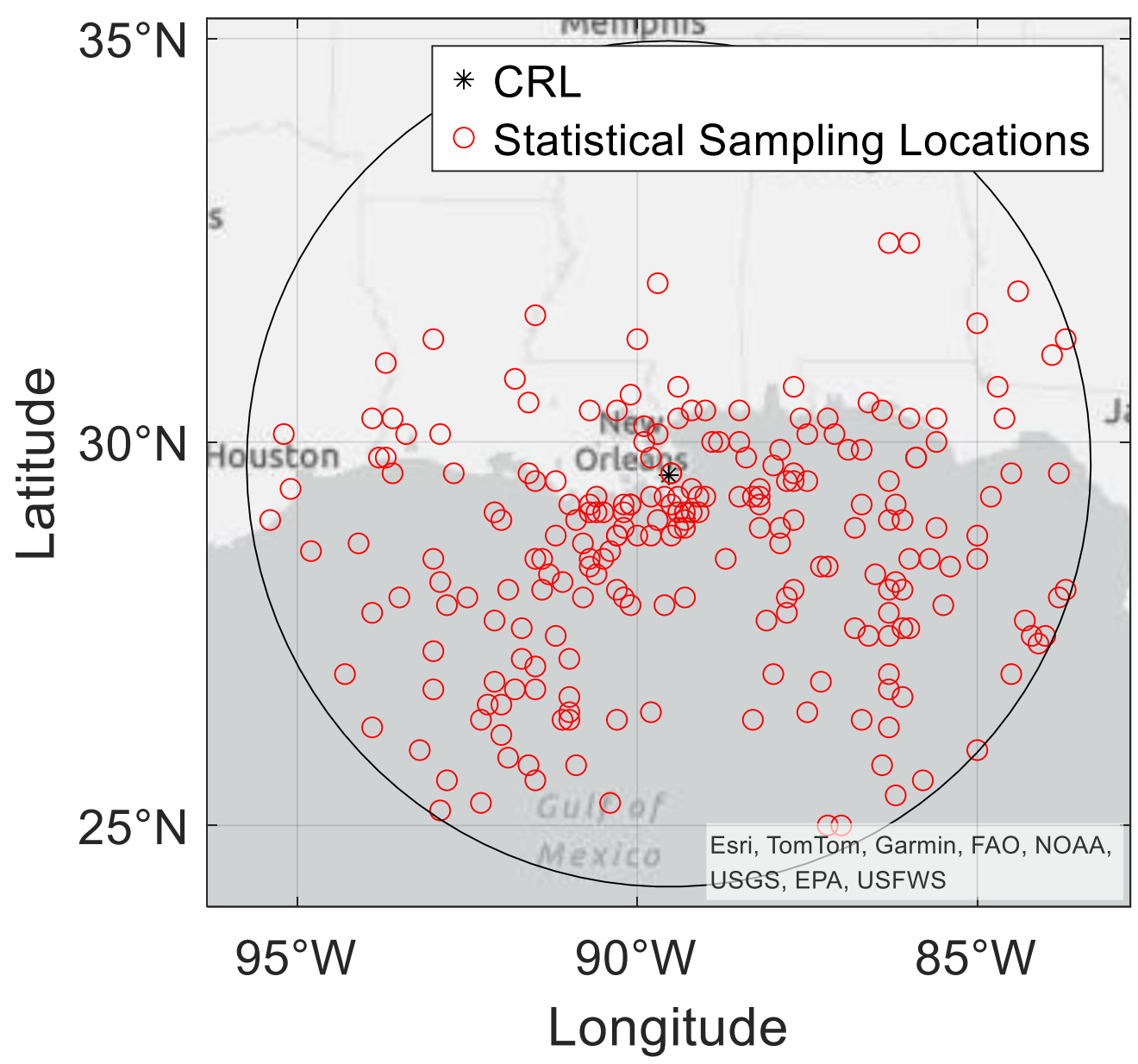

**Figure 9: Locations of CRL, storm statistical sampling data, and capture zone (the black circle)**

Leveraging the work of Nadal-Caraballo et al. (2020, 2022) and Liu et al. (2024b), a meta-Gaussian copula (MGC) is used to construct the joint probability model of TC atmospheric parameters for each season. In this Tier-1 BN, the dependence among TC atmospheric parameters is characterized by separate MGC in each season, with the marginal distributions and parameters (i.e., the correlation matrix) estimated using each season's TC atmospheric parameter statistical sampling data. The fitted marginal PDF curves with the corresponding histogram of statistical sampling data of TC atmospheric parameters in each season are shown in Figure 10. The marginal distribution model (distribution functions) for the TC atmospheric parameters used in this study is consistent with Nadal-Caraballo et al. (2022) and Liu et al. (2024b, 2025). The estimated Kendall's tau for each pair of variables are listed in Table 2. While the fitted marginal PDF curves are similar between Season 1 and Season 2, it is noted that the mode of $\Delta P$ in Season 1 is shifted to the left and is higher compared to Season 2, with a smaller variance. This difference can be attributed to the presence of strong storms in September and reduced activity later in the season. Differences in the estimated Kendall's tau values are observed for the pairs of $V_f$-$R_{max}$ and $V_f$-$\Theta$ between Season 1 and Season 2. The CPT of each TC parameter node is computed using the same method as described by Liu et al. (2025). The discretization information for TC parameters can be found in Appendix A.

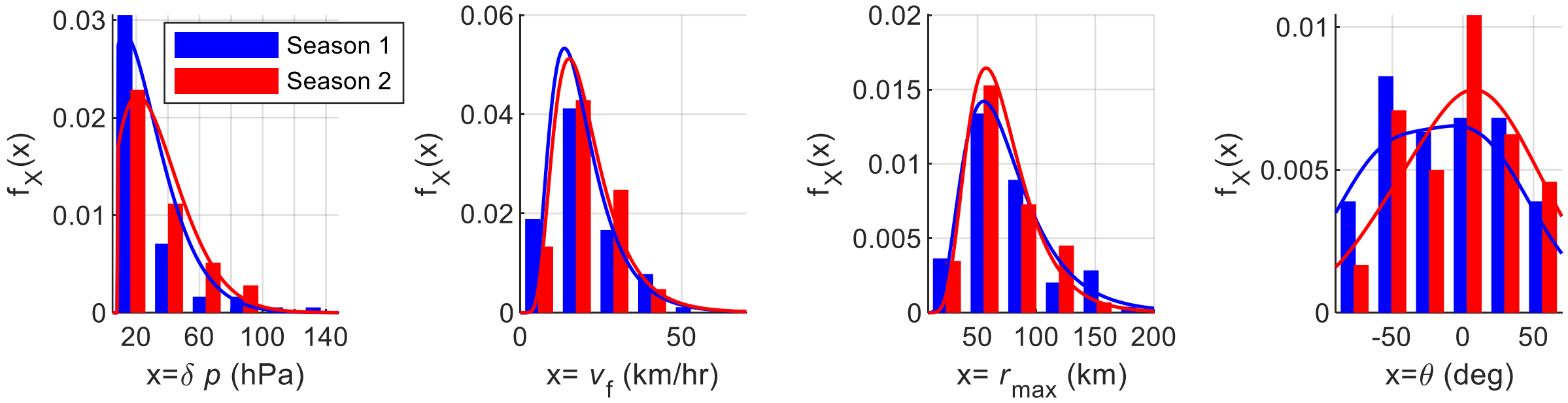

**Figure 10: Fitted marginal PDF curves with the corresponding histogram of statistical sampling data of TC atmospheric parameters; Note scales of y-axis are different in each plot**

**Table 2. Estimated Kendall's tau of seasonality-dependent joint probability models for TC atmospheric parameters**

| | $\Delta P$-$V_f$ | $\Delta P$-$R_{max}$ | $\Delta P$-$\Theta$ | $V_f$-$R_{max}$ | $V_f$-$\Theta$ | $R_{max}$-$\Theta$ |
|---|---|---|---|---|---|---|
| **Season 1** | 0.06 | -0.30 | -0.04 | -0.19 | 0.27 | -0.05 |
| **Season 2** | 0.08 | -0.26 | -0.02 | -0.02 | 0.04 | 0.02 |

## 2.3. Hydrodynamic-model-derived water surface level model

To compute the CPT of node $\hat{Z}$ in the Tier-1 BN, a CCF WSL model incorporating both storm surges and river flow is used. In the past decades, different models have been applied to coastal hazard analysis. A series of USACE studies leveraged ADCIRC and wave models for storm surge simulations (Gonzalez et al. 2019; Nadal-Caraballo et al. 2015, 2020, 2022b). Meanwhile, surrogate models based on machine learning have been well-studied for efficiently predicting WSL based on ADCIRC-derived training data. For example, Jia and Taflanidis (2013) implemented a Gaussian process regression (GPR) model for storm surge prediction with a series of follow-up studies expanding and augmenting the use of GPR (Jia et al. 2016; Taflanidis et al. 2014; Zhang et al. 2018) or leveraging and comparing alternate approaches (Al Kajbaf and Bensi 2020; Gharehtoragh and Johnson 2024; Lee et al. 2021; Liu et al. 2024a). Nadal-Caraballo et al. (2020) added GPR-based metamodels (GPM) to the USACE's CHS-PF for WSL prediction.

A GPR-based metamodel (GPM) developed in Nadal-Caraballo et al. (2022) is used here to predict WSL driven by storm surges and river flow (i.e., to develop the CPT for node $\hat{Z}$ given combinations of TC parameters). The surrogate model training data is derived from ADCIRC simulations of 1145 synthetic storm events that cover multiple river flow scenarios with varying storm parameters. The resulting surrogate model can be used to predict peak WSL using TC parameters and river flow as input. The uncertainty information of this WSL model is represented by the normalized residual node $\varepsilon_Z$, which accounts for both the relative and absolute uncertainty of the modeled value. The process of computing combined uncertainty is described in the Section 6.3 of Nadal-Caraballo et al. (2022). The actual WSL is represented by the node $Z$, which is defined via the superposition of estimated WSL and an error term:

$$Z = \hat{Z} + \varepsilon_Z \sigma_{\mathrm{c}}(\hat{Z}) \tag{4}$$

where $\sigma_{\mathrm{c}}$ is the standard deviation of the combined uncertainty term, which is a function of $\hat{Z}$.

Monte Carlo simulation (MCS) is implemented for the computation of the CPTs of nodes $\hat{Z}$ and $Z$ for the purpose of reducing the error caused by discretization (Mohammadi et al. 2023). In the MCS process, a large number of TC parameter values are generated for each combination of TC parameters' discretized bins and used to define the corresponding conditional distribution of $Z$. For the CPT of node $Z$, a large number of normalized residual values corresponding to each state of $\varepsilon_Z$ are generated to compute the error term and superimposed with simulated values from the discretized states of the node $\hat{Z}$. Similar MCS strategies are applied to other BN nodes in this study to reduce the error induced by discretization. A graphical explanation of this MCS in BN can be found in Mohammadi et al. (2023).

In the node $Z_{\mathrm{t}}$, the tides component is accounted for by superimposing WSL and random generalized astronomical tides. Similar to Equation (4), $Z_{\mathrm{t}}$ is calculated as:

$$Z_{\mathrm{t}} = Z + T\sigma_{\mathrm{t}} \tag{5}$$

where $\sigma_{\mathrm{t}}$ is the standard deviation of astronomical tides computed in the CHS-LA study (Nadal-Caraballo et al. 2022).

## 3. Tier-2 Bayesian Network Construction

In this section, a Tier-2 BN model is constructed for the case study region. In this Tier-2 BN model, the river flow seasonality is preserved to capture base flow ($Q$) seasonality, while a TC intensity node, $I$, is used to categorize TCs into different intensities. TC-induced rainfall is included in the relationship between TC parameters and WSL ($\hat{Z}$) because the rainfall is primarily driven by a model with the same atmospheric parameters as input. A surrogate model derived from a regional hydraulic model (HEC-RAS) is used to simulate the resultant WSL involving storm surges, TC-induced rainfall, and river flow. Figure 11 presents this Tier-2 BN, and Table 3 lists the physical definition of each node.

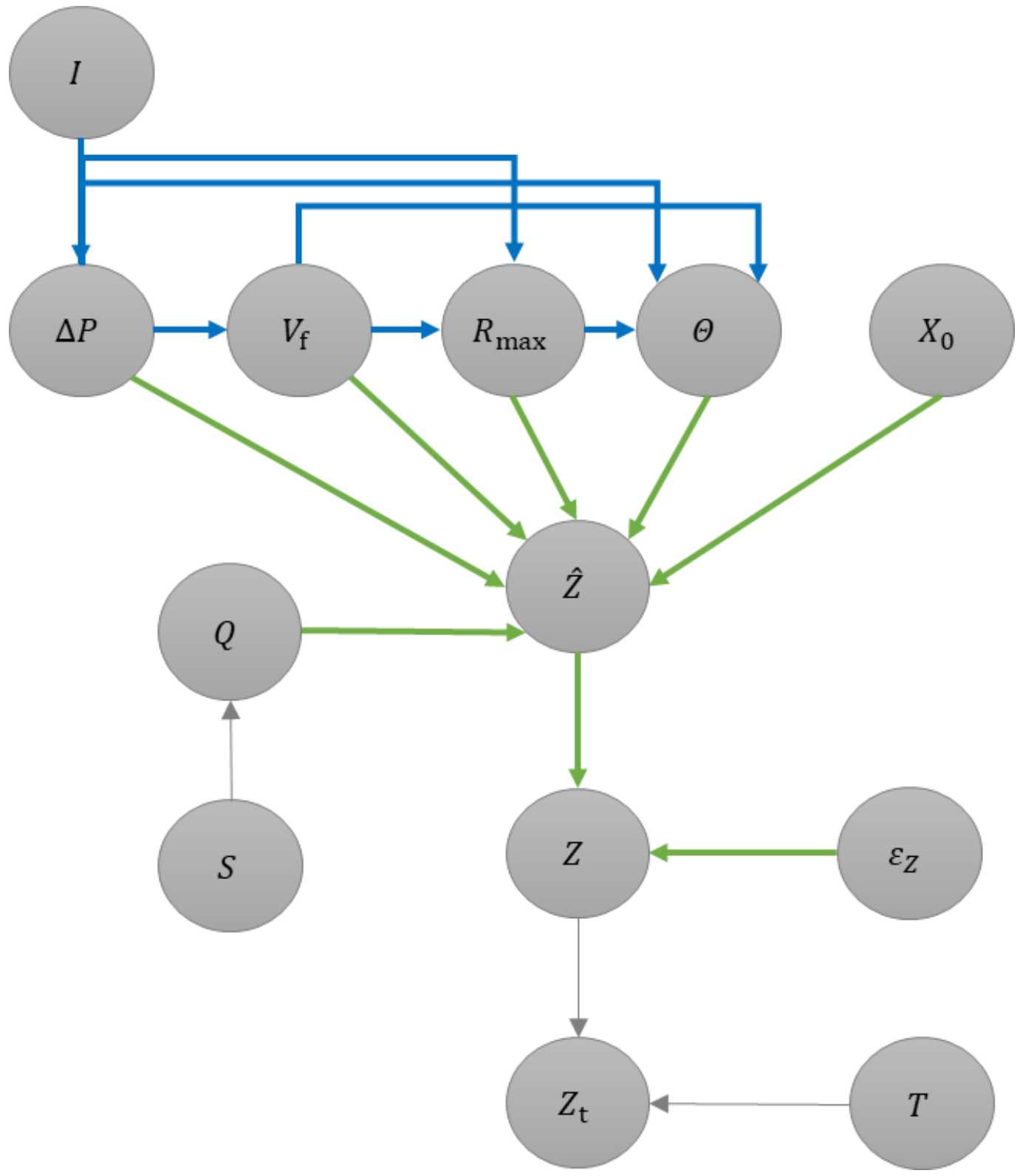

**Figure 11: Tier-2 BN**

**Table 3. Physical definition of each node in Tier-2 BN**

| Notation | Physical definition | Notation | Physical definition |
|---|---|---|---|
| $I$: | Intensity category of TC (LI, MI, HI) | $Z$: | WSL (m) resulting from storm surges, TC-induced rainfall, and river flow with model error considered |
| $\hat{Z}$: | WSL (m) resultant from storm surges, TC-induced rainfall, and river flow predicted by model | | |

Note: Nodes not included have the same physical definition as the Tier-1 BN (see Table 1).

The Tier-2 BN contains three models:

- **Intensity-dependent joint probability model for TC atmospheric parameters** (as identified by blue links in Figure 11): The dependence between TC atmospheric parameters nodes $\Delta P$, $V_{\mathrm{f}}$, $R_{\max}$, and $\Theta$ (for a given TC intensity category) is reflected by the links connecting them, with different distributions specified for each intensity.

- **TC-induced rainfall model**: A rainfall model is used to simulate TC-induced rainfall. The rainfall model is not directly reflected in the BN. Instead, it is used to provide TC-induced rainfall input for the hydraulic model-derived WSL model as a function of the TC parameters.
- **Hydraulic model-derived WSL model** (as identified by green links in Figure 11): The estimated CCF WSL resulting from storm surges, TC-induced rainfall, and river flow is represented by node $\hat{Z}$, which is a child of nodes $\Delta P$, $V_{\mathrm{f}}$, $R_{\max}$, $\Theta$, $X_0$, and $Q$.

Additional details regarding the generation of CPTs associated with each of the above models are provided in the subsections that follow.

## 3.1. Intensity-dependent joint probability model for TC atmospheric parameters

In this Tier-2 BN, the statistical sampling data of TC atmospheric parameters are partitioned based on the value of $\Delta P$ into three categories: low-intensity (LI, 8 hPa $\leq \Delta P < 28$ hPa), moderate-intensity (MI, 28 hPa $\leq \Delta P < 48$ ), and high-intensity (HI, 48 hPa $\leq \Delta P < 148$ hPa). For each TC intensity category, a separate set of marginal distributions is fitted to the intensity-partitioned statistical sampling data of $\Delta P$, $V_{\mathrm{f}}$, $R_{\max}$ and $\Theta$, and a separate MGC is fitted using the procedure described in Liu et al. (2024b).

## 3.2. Tropical cyclone-induced rainfall model

To incorporate TC-induced rainfall in this Tier-2 BN, a rainfall model is used to generate rainfall under synthetic storm events (i.e., to estimate TC rainfall for each combination of TC parameters). In this study, the rainfall induced by synthetic storms is simulated using a tropical cyclone rainfall model (TCR) developed by Zhu et al. (2013) and described in Lu et al. (2018) and Gonzalez et al. (2023). The TCR is a physical-based TC-induced rainfall model and is capable of reliably simulating gridded time series TC rainfall (Lu et al. 2018; Xi et al. 2020; Zhu et al. 2013). The computed rainfall time series is later used as regional rainfall input to HEC-RAS (the USACE Hydrologic Engineering Center's River Analysis System software (USACE 2023)) for CCF modeling, as described below.

## 3.3. Hydraulic model-derived water surface level model

To compute the CPT of the node $\hat{Z}$ in the Tier-2 BN, a CCF WSL model that includes TC rainfall is required. HEC-RAS is a widely used software application that can perform one and two-dimensional hydraulic calculations with rainfall. In this Tier-2 BN model construction, a HEC-RAS 2D regional model adapted from the USACE developed South Louisiana Master Model (Agnew and Marshall 2023) is used to simulate CCFs under storm surges (modeled as a time series boundary), TC-induced rainfall (time series gridded precipitation), and river flow (baseflow at the upstream boundary and rainfall-induced runoff). Figure 12 presents this HEC-RAS 2D regional model. While the model is based on a previously developed model, the adapted model used herein has not been separately validated. A CCF surrogate model is then trained using the HEC-RAS WSL response. The resulting compound model provides a representative "mapping" between TC parameters, base flow, and WSL, which allows the construction of the CPT for node $\hat{Z}$ in this Tier-2 BN.

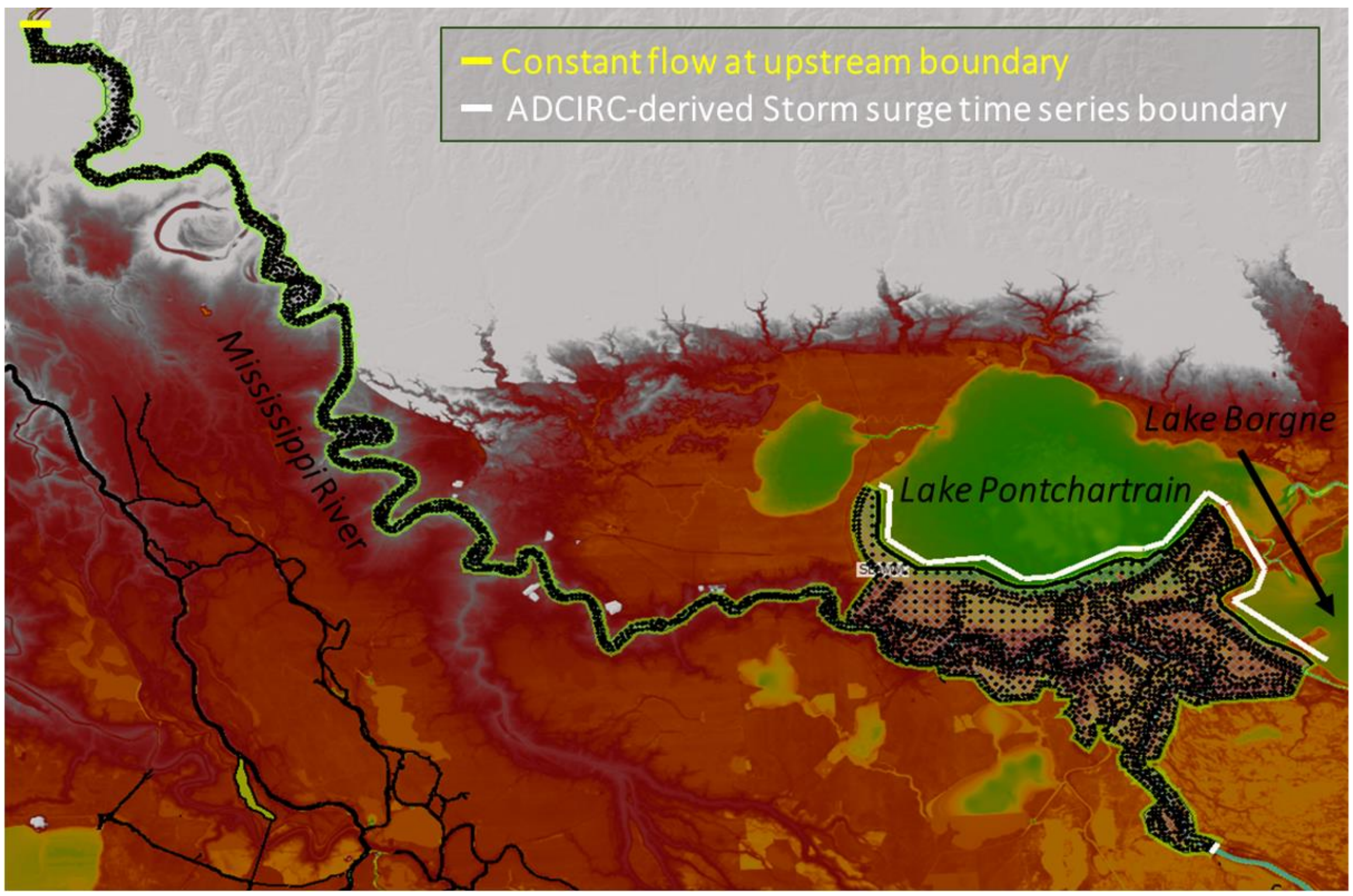

**Figure 12: HEC-RAS 2D regional model; White curves depict boundary condition lines**

This HEC-RAS 2D regional model covers the MR reach starting from around [30.96, -91.66] and ending at a downstream site around [29.63, -89.93]. It includes the region of the Hurricane & Storm Damage Risk Reduction System (HSDRRS) (IPET 2009). The perimeter of the covered region of this model is set based on the contour of the watershed with USGS Hydrologic Unit Code 080701 ("Lower Mississippi-Baton Rouge", a narrow watershed that is mostly levee-restricted) and the HSDRRS protected area.

The HEC-RAS software enables users to incorporate hydraulic structure models into a 2D flow area to simulate the hydraulic mechanics of various types of hydraulic structures that can affect nearby WSL distribution. This is especially relevant for an urban area with a comprehensive flood risk reduction system like the HSDRRS region. In this HEC-RAS 2D regional model, a series of pump stations, flood gates, and barriers are modeled inside the HSDRRS protected area.

As shown in Figure 12, a constant flow is used as the upstream boundary condition of the MR, and ADCIRC storm surge time series are used as a downstream boundary condition for the MR and as boundaries along the coast of Lake Pontchartrain and Lake Borgne. TCR model-derived grided rainfall time series, as mentioned in Section 3.2, is used to cover the whole model area to facilitate TC-induced rainfall input.

A synthetic TC suite adapted from the CHS-LA study (Nadal-Caraballo et al. 2022) is used to generate the aforementioned ADCIRC storm surge and TCR model gridded rainfall time series. A MATLAB program adapted from Leon and Goodell (2016) is developed to automatically input boundary conditions and collect the HEC-RAS simulation result (e.g., WSL) of each synthetic TC event, resulting from the boundary conditions and inputs from storm surges and rainfall. This provided a set of synthetic results that "maps" TC parameters predictors ($\Delta P$, $V_{\mathrm{f}}$, $R_{\max}$, $\Theta$, latitude, and longitude) to peak CCF WSL response. Then, given the demonstrated strong performance of GPR prediction models in storm surge applications (e.g., Jia and Taflanidis 2013; Jia et al. 2016; Taflanidis et al. 2014; Zhang et al. 2018), a GPR surrogate model was trained using the simulation results and then used to compute the CPT of node $\hat{Z}$ in conjunction with the MCS approach as described in the Tier-1 BN. Figure 13 shows the 10-fold cross-validation performance of this surrogate model, which demonstrates the accuracy of this GPR surrogate model.

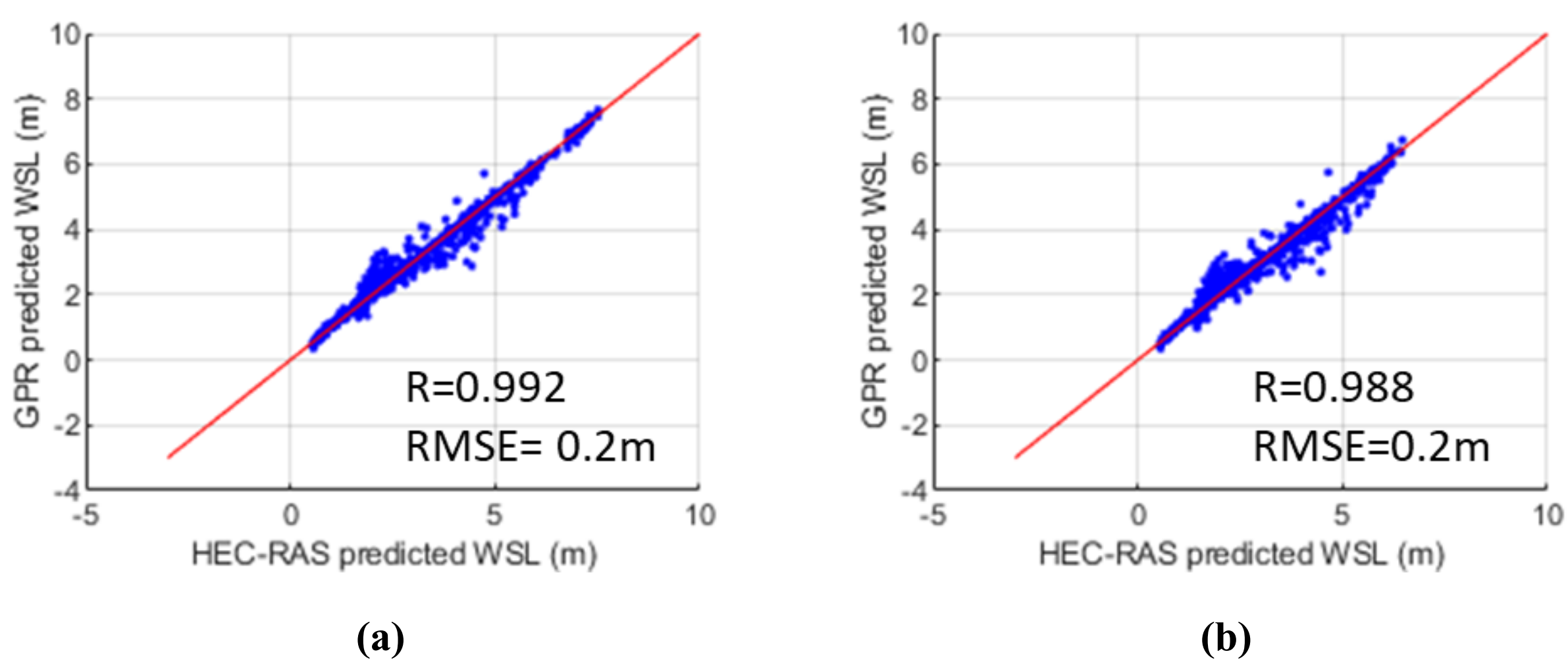

**Figure 13: 10-fold cross-validation performance of the surrogate model. (a) Site 1; (b) Site 2**

## 4. Tier-3 Bayesian Network Construction

In the Tier-3 BN, meteorological and flood antecedent conditions are added to build a CCF BN that can better define uncertainties for future planning and design applications. In this study, the assumptions of Kang and Elsner (2015) are employed to build the relationship between variations of meteorological conditions and TC activity. Based on Kang and Elsner (2015), the ratio of HI TC to all intensity TCs increased with increasing sea surface temperatures (SST) due to variations in meteorological conditions, while the storm recurrence rate (SRR) of all intensity TCs decreased. The BN is sufficiently flexible to accommodate these relationships. In addition, the Tier-3 BN incorporates expanded models for antecedent conditions, namely seasonality of the river baseflow and antecedent precipitation over the river basin. Antecedent coastal water level increases under different meteorological conditions are included in the Tier-3 BN. Figure 14 presents the Tier-3 BN constructed in this study, and Table 4 lists the physical definition of each node inside the Tier-3 BN.

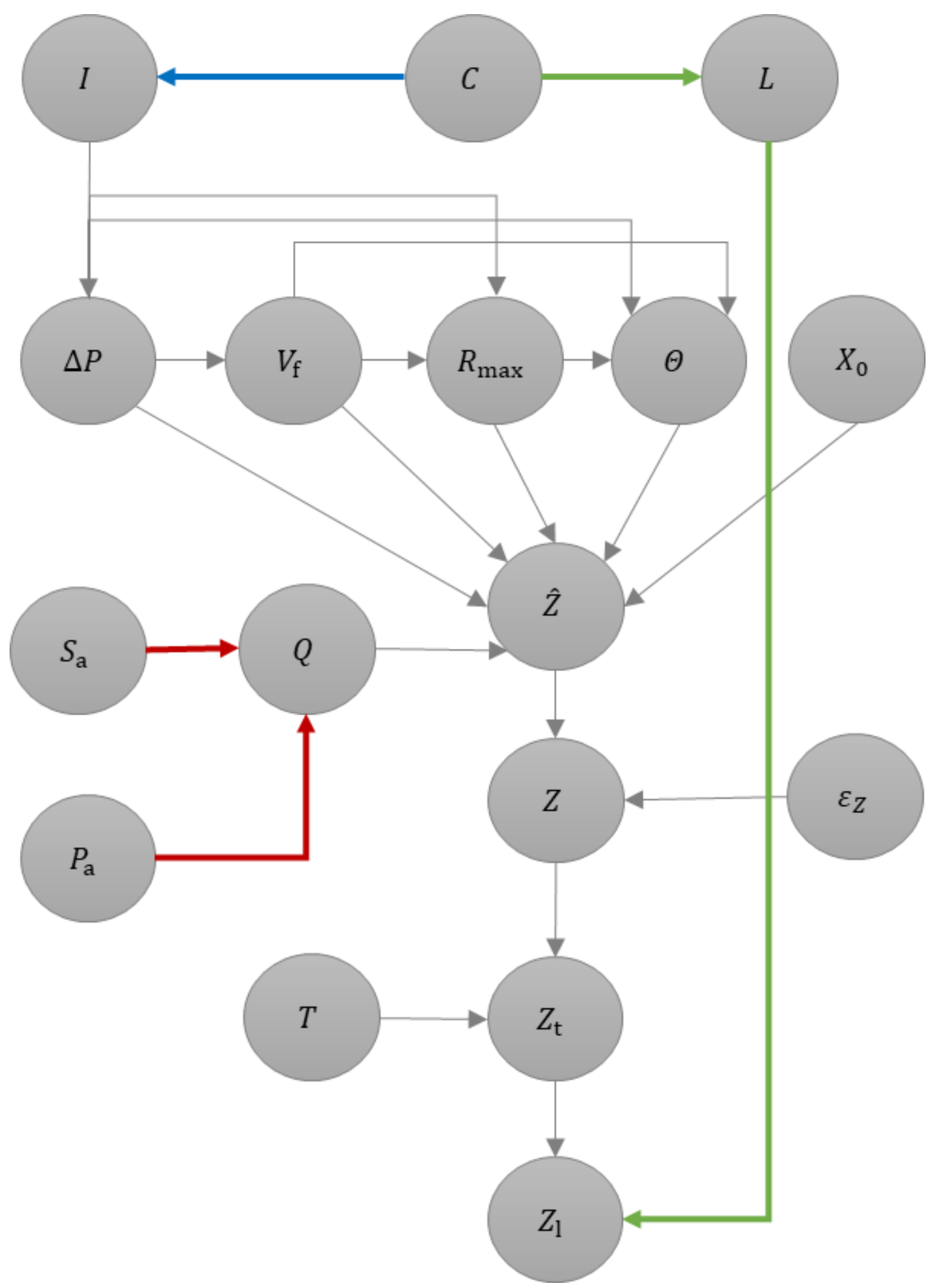

**Figure 14: Tier-3 BN**

**Table 4. Physical meaning of each node in Tier-3 BN**

| Notation | Physical definition | Notation | Physical definition |
|---|---|---|---|
| $C$: | Change of regional SST caused by variations in meteorological conditions | $\hat{Z}$: | WSL (m) resultant from storm surges, TC-induced rainfall, and river flow predicted by model |
| $L$: | Coastal water level increase (m) | $Z$: | WSL (m) resultant from storm surges, TC-induced rainfall, and river flow with model error considered |
| $S_a$: | River season | $Z_l$: | WSL (m) with coastal water level increase involved |
| $P_a$: | Antecedent river basin precipitation (mm) | | |

Note: Nodes not included have the same physical definition as Tier-1 BN and Tier-2 BN (See Table 1 and Table 3).

The Tier-3 BN contains three models:

- **Sea surface temperature-tropical cyclone model** (as identified by the blue link in Figure 14): Potential SST scenarios are captured by node $C$, which is set as a parent node of TC intensity (node $I$) to account for the impact of potential meteorological conditions on TC activity.
- **River antecedent condition model** (as identified by red links in Figure 14): Antecedent condition nodes ($S_a$ and $P_a$) are set as parents of river flow (node $Q$) to account for the impact of antecedent conditions on river flow.
- **Coastal water level increase model** (as identified by green links in Figure 14): Potential SST scenarios are captured by node $C$, which is set as a parent node of coastal water level increase node ($L$) reflecting the impact of meteorological conditions. The final CCF WSL node ($Z_l$) is a child of node $L$ to capture the impact of coastal water level increases on WSL.

Additional details regarding the generation of CPTs associated with each of the above models are provided in the subsections that follow. In addition, the Tier-3 BN leverages models described in the Tier-2 BN. These models are reflected by the grey links in Figure 14.

## 4.1. Sea surface temperature-tropical cyclone model

A series of studies have investigated the impact of variations in meteorological conditions on TC activity. (e.g., Webster et al. (2005), Choi et al. (2019), Knutson et al. (2020, 2013), and Xi et al. (2024)). Because there is currently a lack of well-established consensus models that can predict the characteristics of future TCs, the goal of the Tier-3 BN is to demonstrate the potential of the BN to facilitate the inclusion of meteorological variations on TC-induced coastal storm risk and uncertainties rather than focusing on a specific model.

Based on the findings of Kang and Elsner (2015), this case study implements the assumption that an increase in SST will result in a statistical decrease in the SRR of TCs in the North Atlantic, however, it will also result in higher intensities. Considering the lack of an elaborated SST-TC model and the uncertainty regarding the degree SST increases could impact TCs (Wu et al. 2022), a small range of SST increase is considered in this case study. In the Tier-3 BN, a meteorological condition node ($C$) is used to represent the increase in SST. A simple SST-TC intensity relation is used that assumes that as the SST increases, the SRR decreases linearly. Furthermore, a trade-off is assumed to exist between HI and LI cyclones, i.e., with the increase of SST, the proportion of HI TC increases while the proportion of LI TC decreases. The simplified SST-TC relation is used to build the CPT of the intensity node $I$ in the Tier-3 BN.

## 4.2. River antecedent condition model

Previous studies have investigated the impact of flood antecedent conditions on flood events. For example, Michele and Salvadori (2002) present a peak flood analytical formulation that involves antecedent moisture conditions based on river season and antecedent rainfall. Bennett et al. (2018) investigate the effects of antecedent precipitation on flood volume. Grillakis et al. (2016) investigate the importance of the initial soil moisture state for flash flood magnitudes. In this case study, the river seasonality and antecedent precipitation are combined to build a river antecedent condition model to enable increased refinement in the modeling of antecedent conditions.

Following the assumptions of Michele and Salvadori (2002), two nodes are used to represent antecedent conditions of river in this case study, i.e., river season ($S_a$) and antecedent river basin precipitation ($P_a$ ). The node $S_a$ is defined in the same way that $S$ node is defined in Tier-1 BN. For node $P_a$, the antecedent river basin precipitation value is represented by the basin-averaged accumulated rainfall for the prior five days. PRISM rainfall data (PRISM 2023) is used as a database to inform the development of the CPT for node $P_a$. Specifically, five-day accumulated basin-averaged rainfall data from 1980~2015 are randomly selected to generate independent statistical samples for $P_a$ and a generalized extreme value distribution (MathWorks 2022) is used for marginal distribution fitting. A linear regression rainfall–runoff model is derived from HEC-RAS rainfall–runoff simulations. River flow (node $Q$) is modeled as the superposition of seasonal base flow and runoff induced by antecedent river basin precipitation (node $P_a$). This linear regression rainfall-runoff model performance is reflected in Figure 15. It can be observed that the linear regression rainfall-runoff model has a reliable performance in this case study.

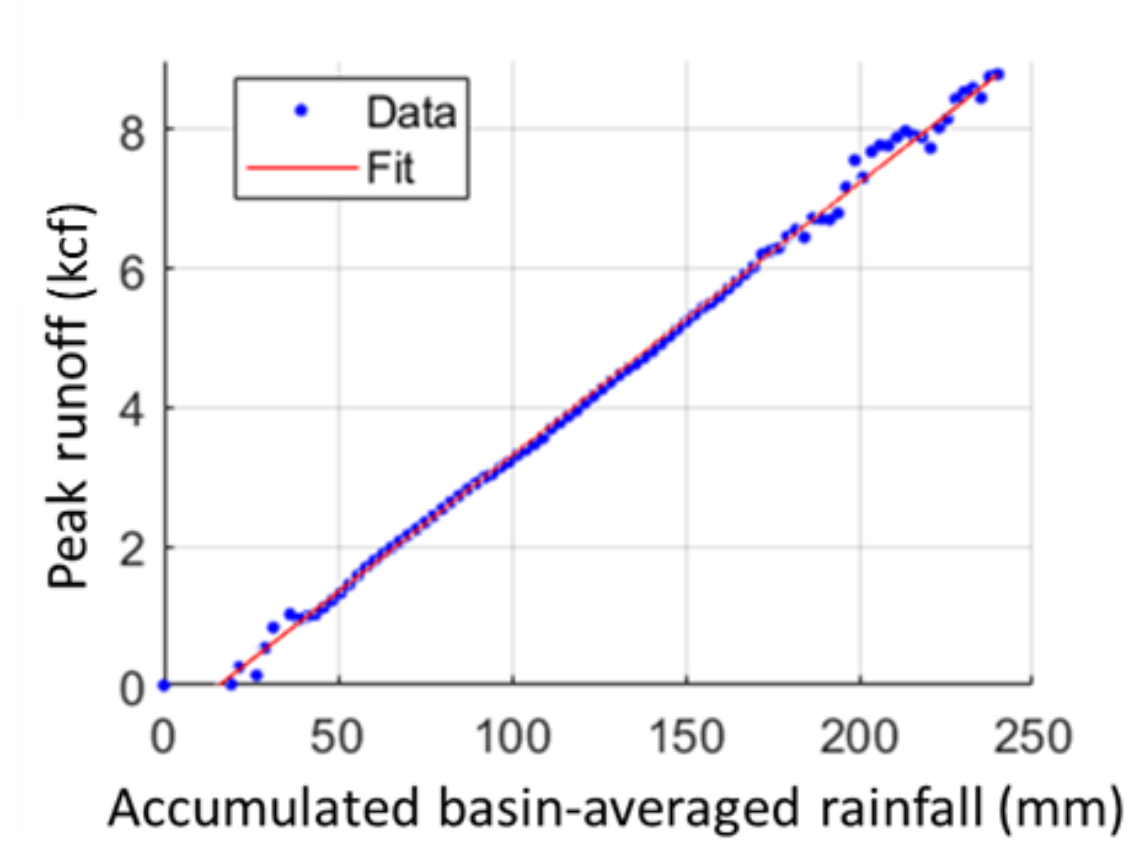

**Figure 15: Linear regression rainfall-runoff model**

## 4.3. Coastal water level increase model

In this case study, a node representing the coastal water level increase (node $L$) is added as a child of the SST scenario (node $C$). An empirical cubic function from Roper (2016) is used to characterize the relationship between SST and coastal water level increase to define the CPT between node $C$ and node $L$. Kyprioti et al. (2021) investigated the performance of incorporating coastal water level increase scenarios into storm surge surrogate models. In the North Atlantic Coast Comprehensive Study (NACCS) (Nadal-Caraballo et al. 2015), the linear superposition method is used to incorporate coastal water level increase into the water level results due to the complexity and extent of the hydrodynamic model. It is noted that the NACCS discusses the potential error induced by linear superposition. In this study, for simplicity, coastal water level increases are superimposed on WSL (i.e., $Z_l = Z_t + L$) as a means of generating the CPT for node $Z_l$.

## 5. Results

As described in Section 2 ~ Section 4, three tiers of BN were constructed for a case study in New Orleans, LA. Leveraging these three BNs, probabilistic analyses of CCF are performed. In Section 5.1, the CCF WSL distributions generated from different BNs are presented to demonstrate the capabilities of the proposed BNs. It is emphasized that the BNs are built using illustrative, limited-scope models and statistical analysis results that differ from other studies. Thus, the results may not reflect actual hazards at the study sites. Nevertheless, the results presented in Section 5.1 demonstrate the benefits and challenges associated with using the proposed tiered framework for assessing coastal storm hazards and risk for planning and design applications. As a supplemental benefit of the use of BNs, Section 5.2 shows how this proposed framework can leverage a BN-based deaggregation method developed by Liu et al. (2025) to better assess CCF hazards.

## 5.1. Coastal compound flood distribution

The distributions of CCF WSL under two seasonalities are generated using the Tier-1 BN, where the node $S$ is set to Season 1 and Season 2 in turn to produce distributions under the two seasonalities separately. The action of setting nodes to particular states is also referred to as entering evidence in BN analysis. Figure 16 plots the CCF WSL distributions.

Distributions are only presented for Site 1 because of the small difference between Site 1 and Site 2 due to the short geospatial distance between the two sites. Figures plotting the results of both sites can be found in Appendix B. It can be observed that the CCF WSL distribution under Season 1 has a slightly higher modal WSL value than Season 2, which is attributed to the seasonal differences in river flow and storm parameter probabilities.

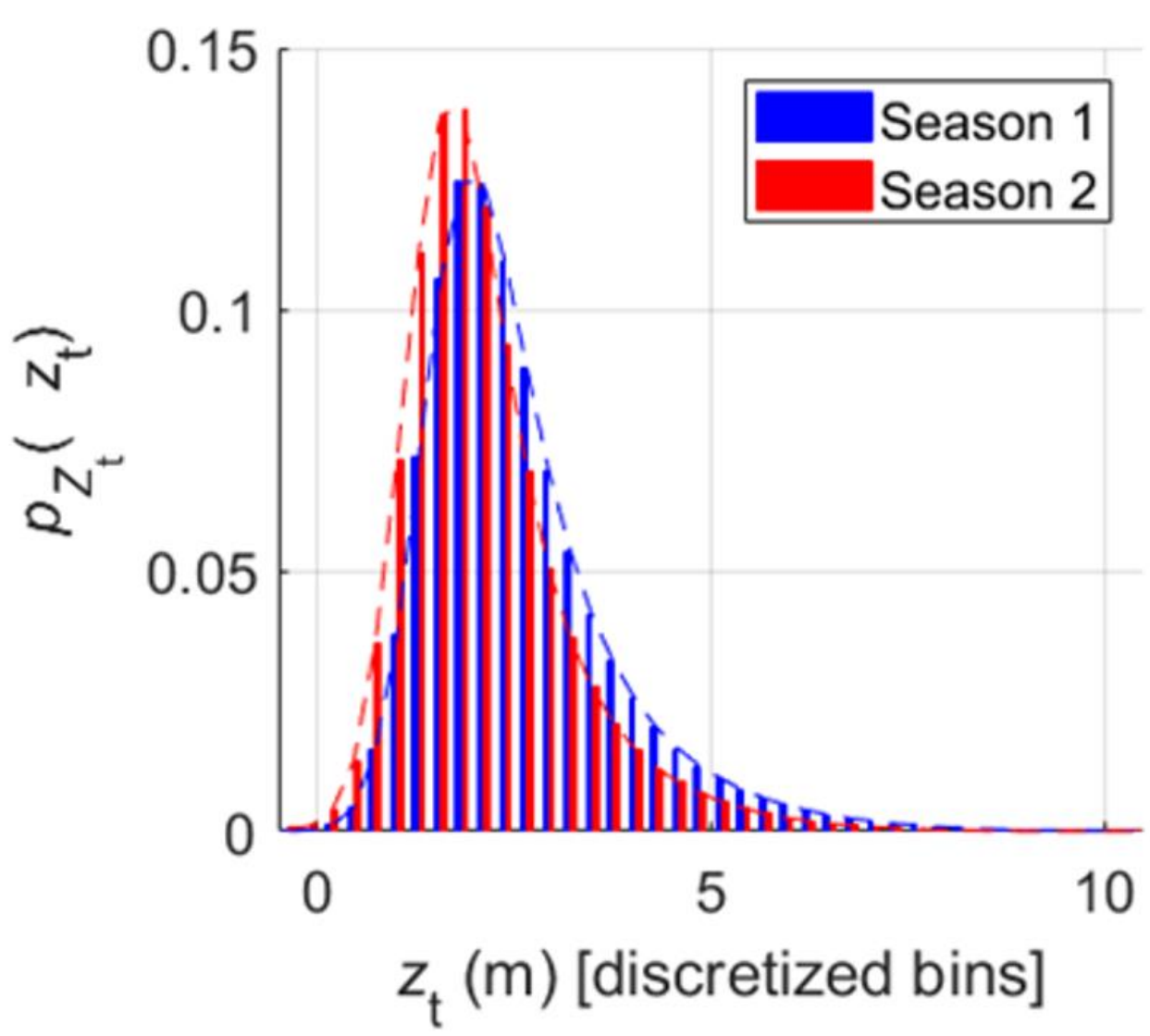

**Figure 16: CCF WSL ($Z_t$) distribution generated by Tier-1 BN at Site 1 (Site 2 not shown due to minor differences in results)**

Figure 17 plots the CCF WSL distributions under three TC intensity categories, generated by the Tier-2 BN through entering evidence into the node $I$.It can be observed that CCF WSL increases with TC intensity (i.e., the probability distributions shift rightward with increased variance), which reflects the connection between higher ΔP TCs and higher resultant CCF WSL, as well as greater uncertainty in the resultant CCF WSL.

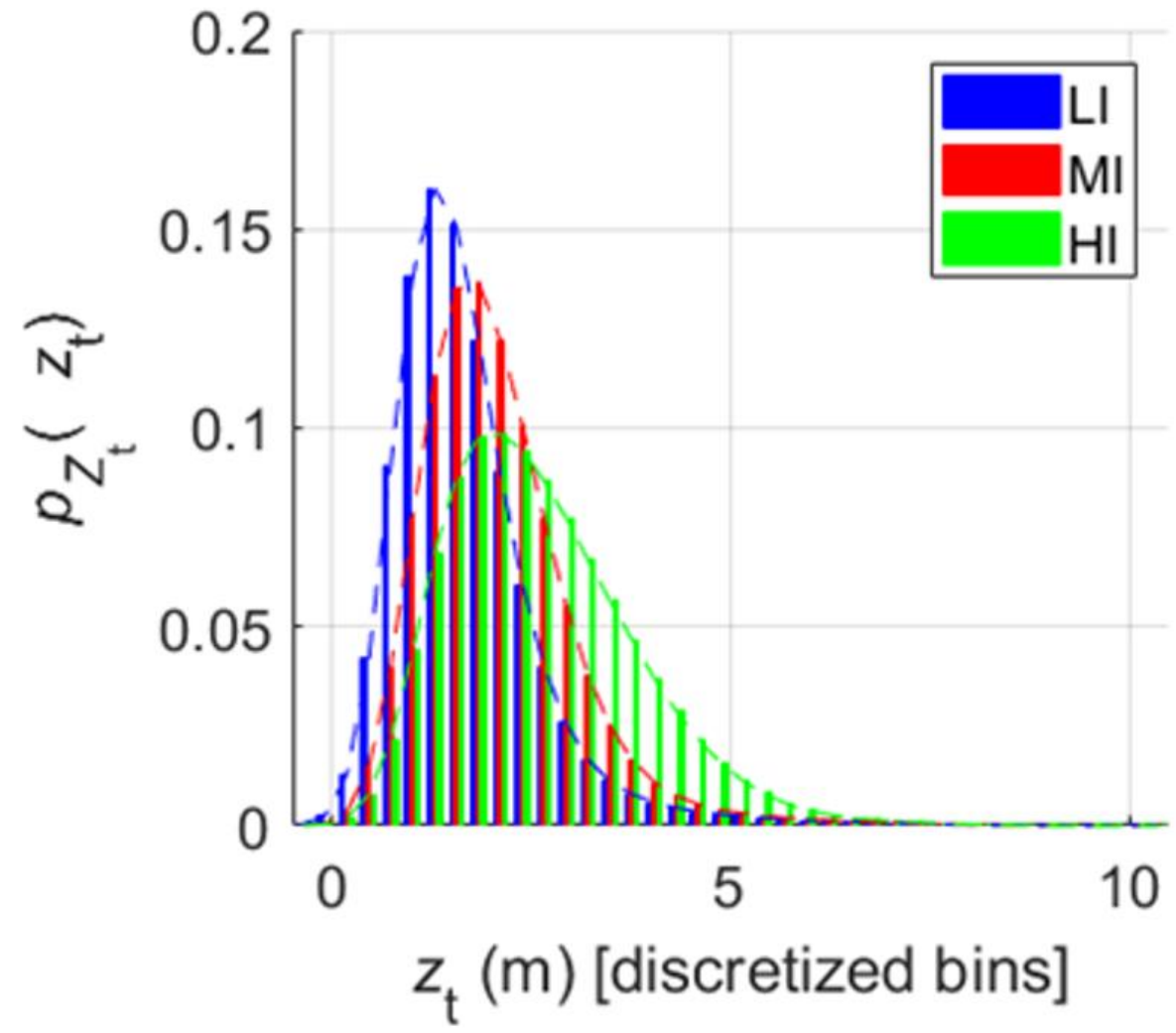

**Figure 17: CCF WSL ($Z_t$) distribution generated by Tier-2 BN at Site 1 (as above)**

Figure 18 plots the CCF WSL distributions and hazard curves under illustrative SST increase scenarios for both study sites, generated by the Tier-3 BN through entering evidence into the node $C$. A clear trend can be observed that with increases of SST, the CCF WSL distributions shift rightward, which represents a more severe CCF hazard. Meanwhile, with increases of SST, the predicted hazard curves shift up (hazard curves reflect reduced SRR under the increase of SST). In general, when comparing Site 1 and Site 2, it can be observed that Site 2 consistently exhibits a lower hazard curve across all cases. This difference can be due to Site 2 being located downstream of Site 1 along the river, allowing it to benefit more from hydraulic structures that release floodwaters into nearby lakes.

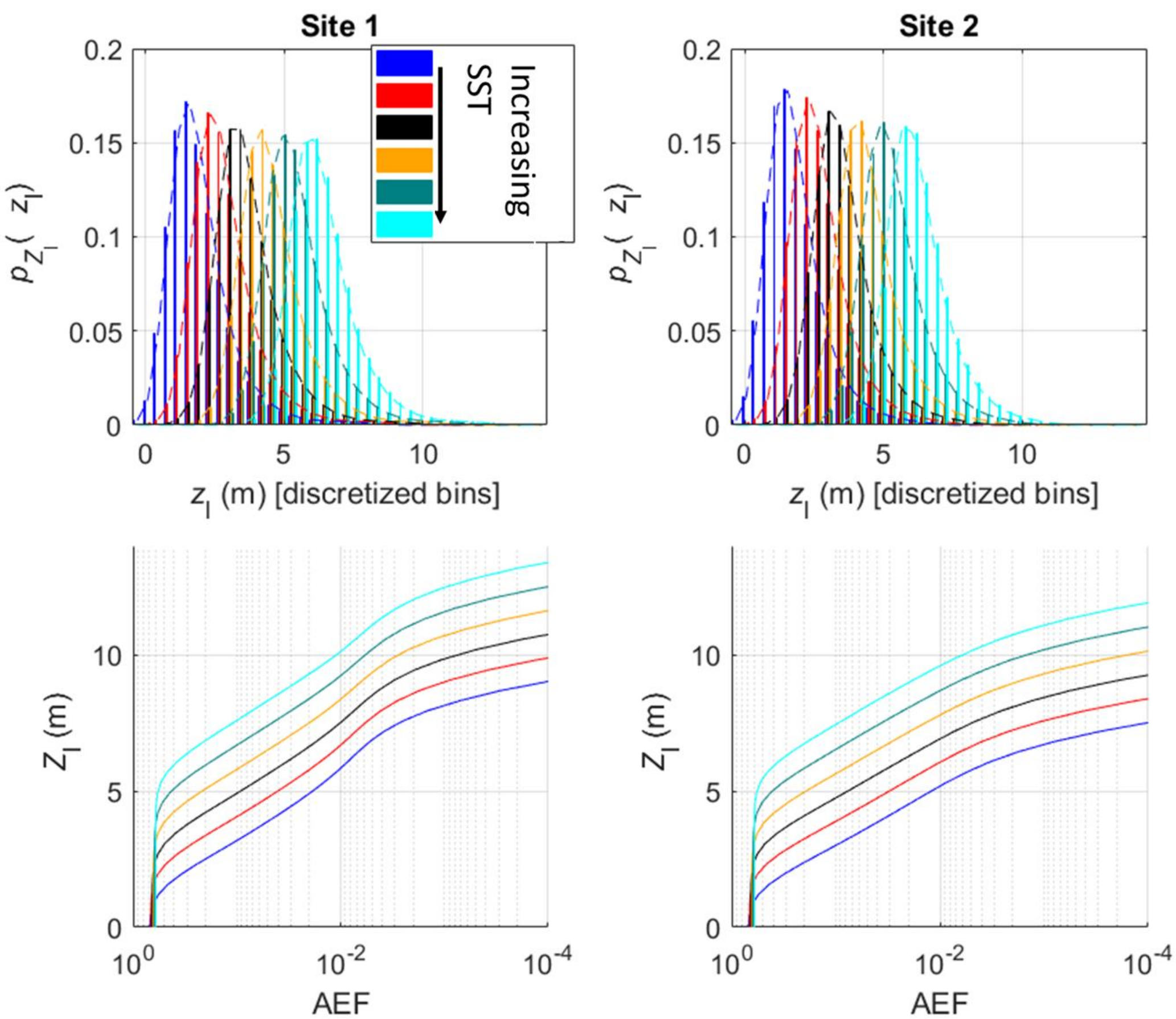

**Figure 18: CCF WSL ($Z_1$) distributions generated from the Tier-3 BN (top row); Representative hazard curves (CCF WSL) generated from the Tier-3 BN (bottom row), the x-axis shows annual exceedance frequency**

Using the representative assumptions and models employed for the case study, Figure 19 plots the CCF WSL hazard curves generated from the Tier-1 BN, Tier-2 BN, Tier-3 BN under a moderate SST increase scenario, and Tier-3 BN under a low SST increase scenario for comparison. It is noted that although the Tier-2 BN incorporates TC-induced rainfall in the CCF simulation, it produces a lower hazard curve than the Tier-1 BN. This result can be primarily explained by the fact that Tier-1 and Tier-2 BNs used different models for WSL simulations (i.e., ADCIRC in the Tier-1 BN and HEC-RAS in the Tier-2 BN). Additionally, the limited impact of rainfall at the two study sites, as well as the differences in the joint distribution models for TC atmospheric parameters between the Tier-1 and Tier-2 BNs, might all contribute to the fact that the Tier-2 BN produces a lower hazard curve than the Tier-1 BN. As expected, the Tier-3 BN under moderate SST increase yields a higher hazard curve than both the Tier-1 and Tier-2 BNs, aligning with the expectation that an increased SST will increase the severity of CCF hazards. The Tier-3 BN under a low SST increase scenario hazard curves are generated by specifying a near-zero SST increase in the Tier-3 BN (i.e., setting node $C$ to the first state with discretized bin [-0.05, 0.05] °$C$). Its relatively minor difference with the Tier-2 BN hazard curves can be attributed to the inclusion of river antecedent conditions and MCS processes in node $C$ and node $L$ within the Tier-3 BN. Again, when generally comparing Site 1 and Site 2, it can be observed that Site 2 exhibits a lower hazard curve under different tiers of BNs, which is consistent with the comment in Figure 18. Nevertheless, it is important to note that the hazard curve results presented here are intended to illustrate the multi-tier BN framework and may differ from results in existing studies in the region due to variations in models and assumptions used. Moreover, observations regarding the relationship between the resulting hazard curves may differ depending on model assumptions.

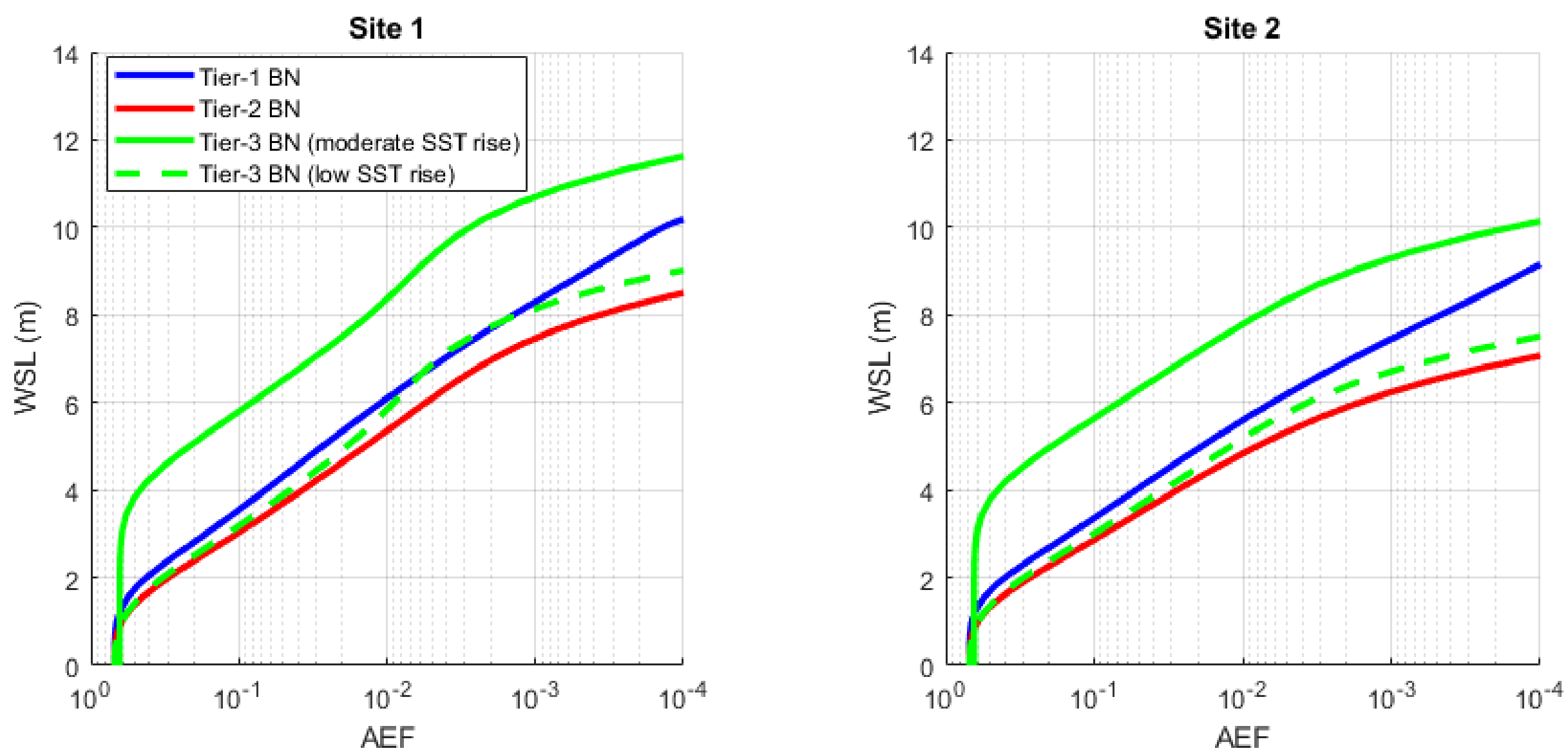

**Figure 19: CCF WSL Hazard curve generated from the Tier-1 BN (node $Z_{\mathrm{t}}$), Tier-2 BN (node $Z_{\mathrm{t}}$), Tier-3 BN under a moderate SST increase (node $Z_{\mathrm{l}}$), and Tier-3 BN under a low SST increase (node $Z_{\mathrm{l}}$)**

## 5.2. Coastal compound flood deaggregation

Liu et al. (2025) have developed a BN-based coastal hazard deaggregation method based on seismic hazard deaggregation methods (Baker 2013; Bazzurro and Allin Cornell 1999). In this deaggregation method, leveraging the BN backward inference analysis, the probability of discretized TC parameters given coastal hazard $R > r$ can be computed:

$$P(\mathbf{X} = \mathbf{x}|R > r) = \frac{P(\mathbf{X} = \mathbf{x} \cap R > r)}{P(R > r)} \tag{6}$$

where $\mathbf{x} = [\delta p, v_{\mathrm{f}}, r_{\max}, \theta, x_0]$ is a combination of discretized TC parameters. In this study, $R$ can be interpreted as the CCF WSL and $r$ can be interpreted as a specified threshold CCF WSL value. This deaggregation method can provide insights regarding which portions of the TC parameter domain are most relevant to severe coastal hazards, with relatively limited computing resources. This is helpful for risk-informing and decision-making processes (e.g., scenario selection) and identifying the potential need for refined CCF events modeling (e.g., high-fidelity numerical simulation).

Here, this deaggregation method is applied. Considering the employed discretization in BNs, a threshold WSL value of 5 m (close to the 100-year CCF WSL value at both Site 1 and Site 2 under the Tier-2 BN) is selected as a representative WSL of severe CCF. An evidence case (EC) (i.e., a case in which the value of a BN node is assigned particular known or assumed states) of CCF WSL > 5 m is entered into node $Z_{\mathrm{t}}$ to demonstrate the capability of this tiered BN framework to identify the posterior distributions of TC parameter combinations given significant CCF events.

First, to visualize the track information (information on TC track discretization can be found in Appendix A) for dominant TCs, Figure 20 shows the pair-wise posterior joint probability mass (PM) plot of $X_0$ (landfall location) and $\Theta$ (track angle) when ECs are entered into the Tier-1 BN, Tier-2 BN, and Tier-3 BN (with a moderate SST increase scenario). The posterior distribution provides information regarding the probability of storm track parameters given observations of severe CCF events ($R > 5$ m). This provides insights regarding the storm tracks most likely to contribute to severe CCF events. The no-evidence case (i.e., considering only the initial assumptions used to build the model prior to entering any evidence in the BN) is plotted for comparison. It can be observed that, in the no-evidence case, the pair-wise joint PM of $X_0$ and $\Theta$ are uniformly distributed along $X_0$. This is because a uniform probability distribution is applied to $X_0$ in this study (See Appendix A).

The input of evidence notably affects the pair-wise joint PM. Across all tiers, the dominant track parameters are concentrated around similar values ($X_0$ ranging around landfall location index 6-8 and $\Theta$ ranging from -50 to 0 degrees).  Compared with the Tier-1 BN, the Tier-2 BN produces results that are generally more consistent with expectations. The Tier-1 BN results include a cluster near the extreme landfall and heading values (in the upper-right corner of the pattern). This may reflect limitations of the models employed in the Tier-1 BN to develop the CPTs. This more coherent pattern in the Tier-2 BN suggests that, under the modeling assumptions employed in the Tier-2 BN, the results are more stable. Under the Tier-3 BN (with a moderate SST increase scenario), a notable expansion of the range of dominant TC tracks compared to the Tier-2 BN is observed. This suggests that, under a moderate SST increase scenario, severe CCFs are more likely to be triggered by TC tracks that would not have triggered severe CCFs without considering SST increase (as compared with the Tier-2 BN).

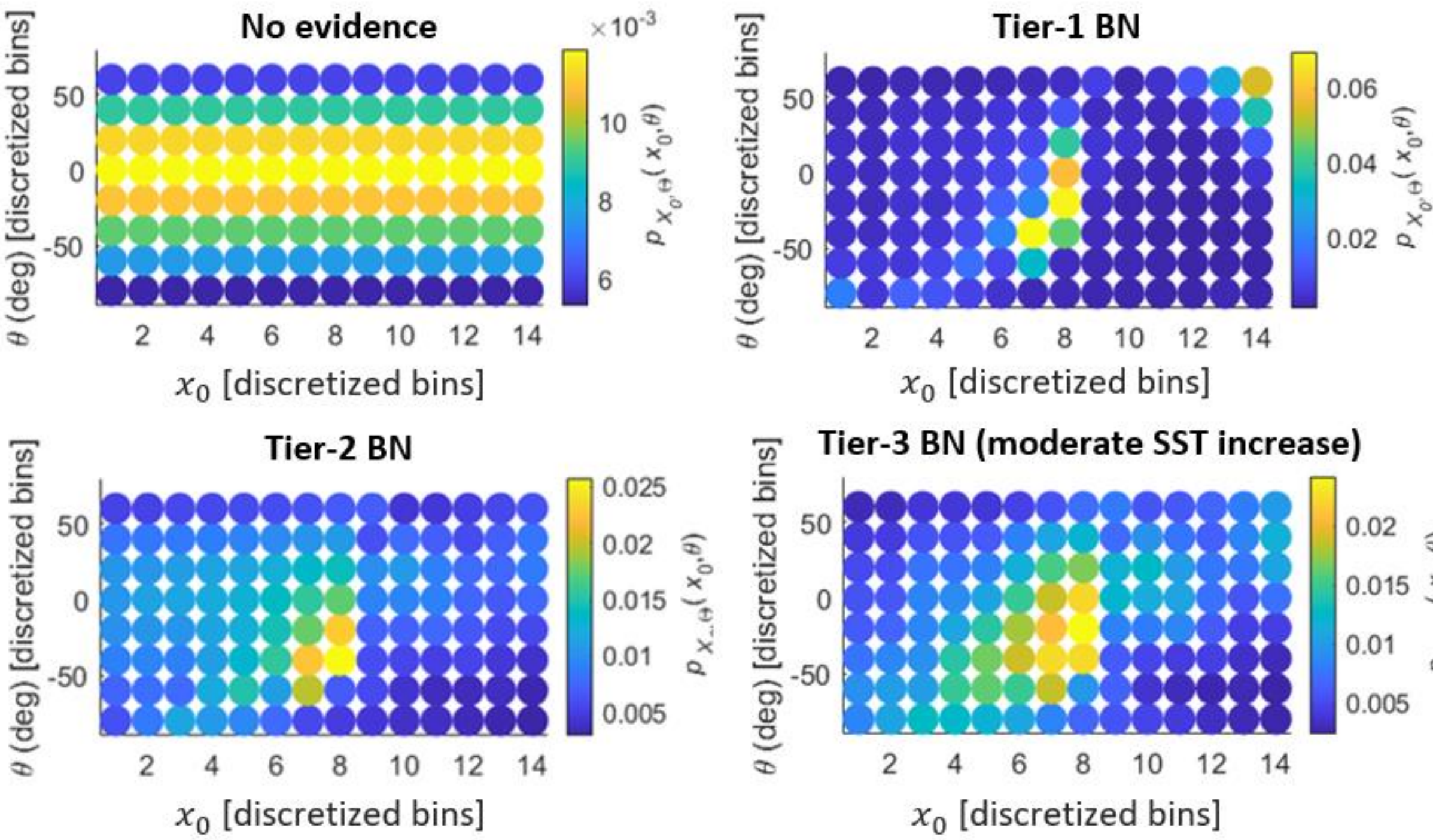

**Figure 20: Pair-wise joint PM of $X_0$ and $\Theta$ under ECs (WSL > 5 m) at Site 1. Note that the range of color bars are different in each plot (Site 2 is not shown but is associated with similar results, results plots of both sites can be found in Appendix B)**

The contributions of $\Delta P$, $V_{\mathrm{f}}$ and $R_{\max}$ in conjunction with the TC track variables ($X_0$ and $\Theta$) are visualized in Figure 21 to Figure 23 as 3D stacked bar plots with partitions identified with different colors. From Figure 21, which focuses on the contribution of central pressure deficit $\Delta P$, it can be observed that TCs with high $\Delta P$ (as colored in yellow) dominate the most likely TC tracks for the high WSL EC. Under the Tier-3 BN (with a moderate SST increase scenario), the portion of relatively low $\Delta P$ slightly increased compared to insights from other tiers of BN. This can be explained by recognizing that, under SST increases-caused coastal water level increases, a relatively low intensity TC event can easily generate severe CCFs.

Figure 22, which focuses on the contribution of forward velocity, $V_{\mathrm{f}}$, shows that moderate $V_{\mathrm{f}}$ (15 km/hr $< V_{\mathrm{f}} \leq 25$ km/hr, colored as cyan) dominates the most likely TC tracks under EC, which is similar to the no-evidence case. Figure 23**,** which focused on the contribution of radius to maximum winds, $R_{\max}$, shows that the contribution of small $R_{\max}$ (10 km $\leq R_{\max} \leq 60$ km, colored as dark blue) dominates most dominant TC tracks under ECs, which is also similar to the no-evidence case.

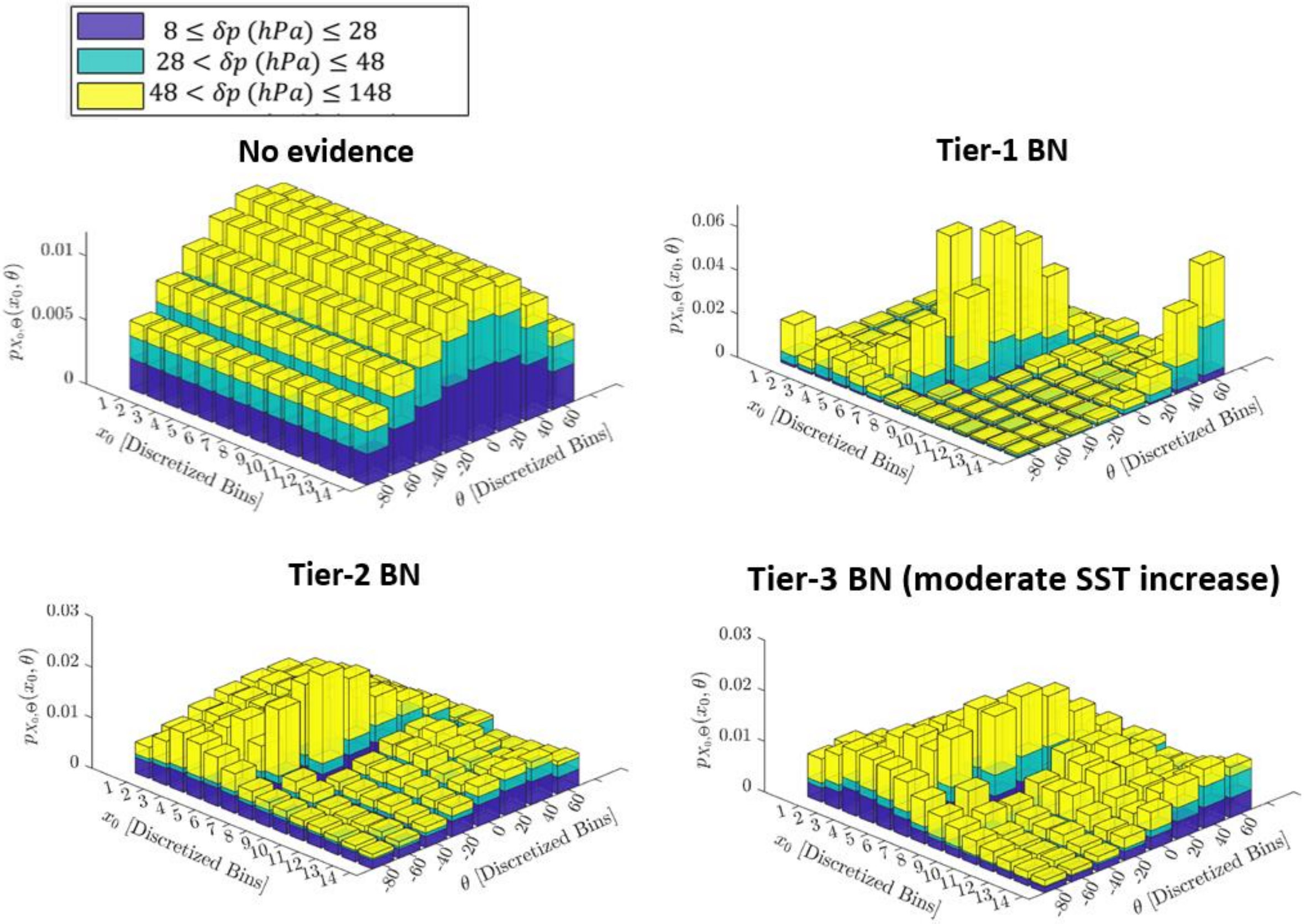

**Figure 21: 3D stacked bar plot of TC track information with $\Delta P$ identified by colors at Site 1**

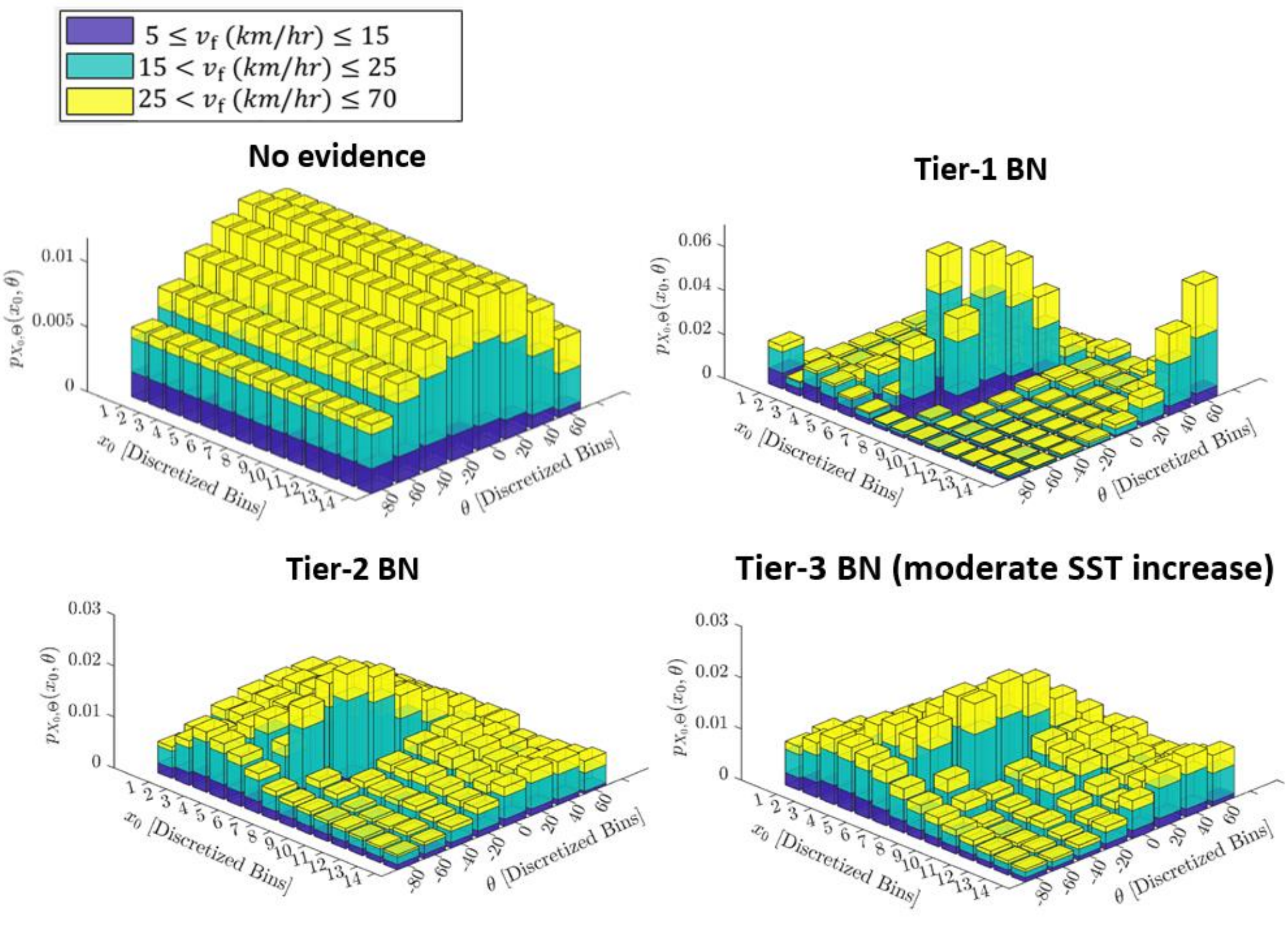

**Figure 22: 3D stacked bar plot of TC track information with $V_f$ identified by colors at Site 1**

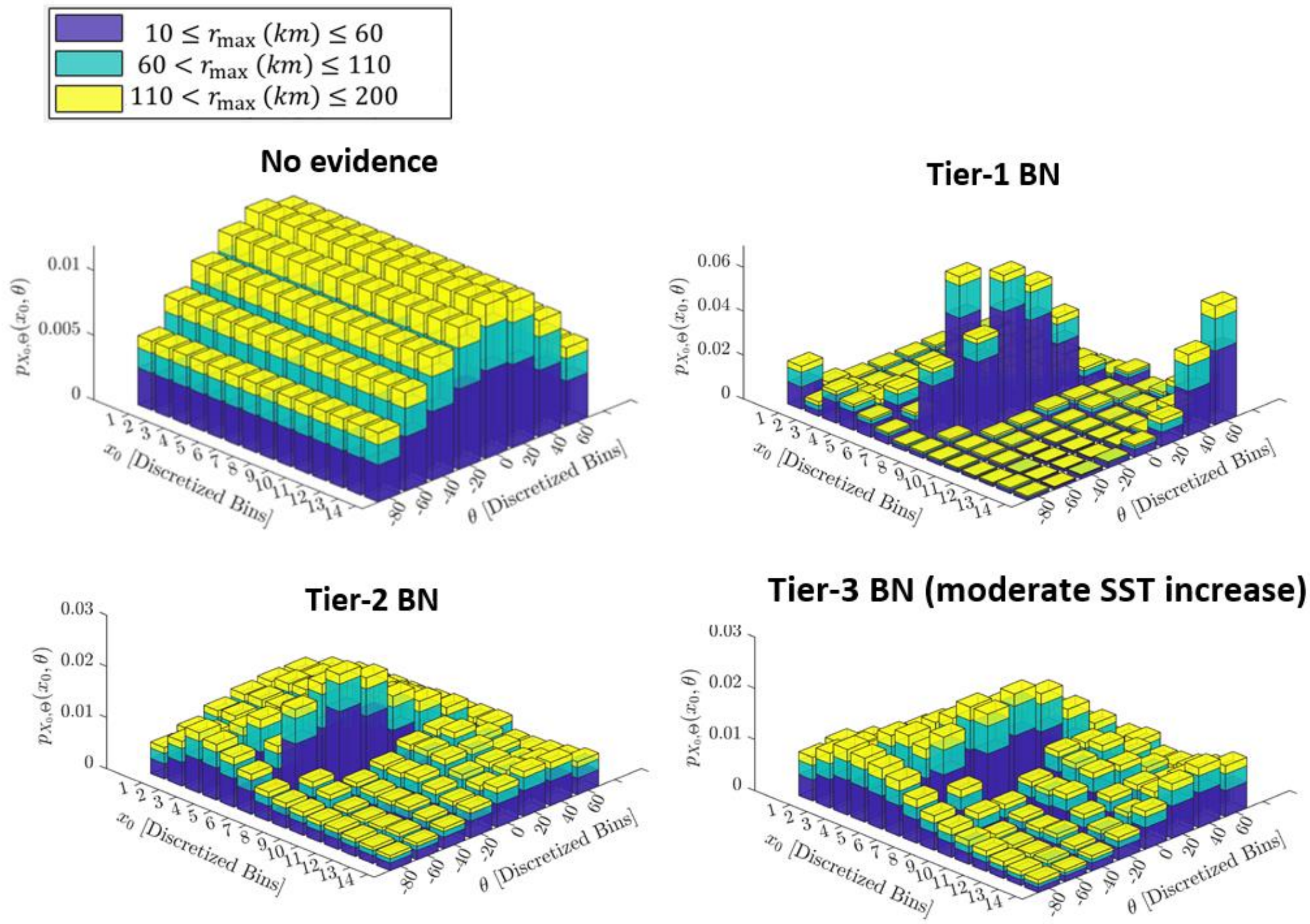

**Figure 23: 3D stacked bar plot of TC track information with $R_{\max}$ identified by colors at Site 1**

## 6. Conclusion and Discussion

In this work, a framework for constructing multiple tiers of CCF BN model is established. The motivation behind this framework lies in the idea of creating tiered BN models that are progressively more sophisticated as more data and computing resources become available or more aspects of the compound coastal hazard need to be represented due to the application. The simplest BN considers only storm surge and river processes as mechanisms contributing to CCF, while the most sophisticated BN incorporates both rainfall and potential variations in meteorological conditions as assessed CCF mechanisms.

The framework is implemented for an illustrative case study conducted in New Orleans, LA. A series of joint distribution, numerical, machine learning, and experimental models are used to build specific multiple tiers of BN for this study region. CCF probabilistic analyses and deaggregations are performed based on the BNs. Although a series of simplifications and coarse assumptions are used herein, the results demonstrate the promise of these BN models in generating CCF hazard assessments under various levels of resource availability and project needs.

The deaggregation results highlight the potential of using this series of BN models for the identification of dominant TC of CCF events that might cause significant damage to communities. This is helpful for both risk-informed decision-making for planning and design and for identifying the potential need for refined CCF events modeling (e.g., high-fidelity numerical simulations). In this illustrative case study, the Tier-2 BN provides deaggregation results that are more consistent with expectations compared to the Tier-1 BN. Additionally, the Tier-3 BN, building on the Tier-2 BN, is capable of capturing the trend of CCF risk with SST increase.

The data and computing resources needed to implement this tiered BN framework may differ depending on the study region. In particular, additional nodes, dependence models, and more sophisticated process models may be necessary to account for the temporal process of CCF mechanisms in a given region (Joyce et al. 2018). In the Tier-3 BN, alternative strategies like linking future scenarios relevant for coastal storm risk management can be explored. As models continue to evolve, there is also room for integrating more sophisticated regional TC activity models into this framework.

In summary, the adaptability and complexity of the BN framework can be tailored to the unique characteristics and data availability of the study region, allowing for a more comprehensive representation of CCF events.

## 7. Acknowledgment

### Data

Historical datasets used in this study are available online. HURDAT2 dataset can be obtained from NOAA at https://www.aoml.noaa.gov/hrd/hurdat/Data_Storm.html. EBTRK dataset can be obtained from CIRA at https://rammb2.cira.colostate.edu/research/tropical-cyclones/tc_extended_best_track_dataset. Numerical coastal and hydraulic model were provided by the developers for use in this study. Original development and runs were led by T. Chris Massey at USACE/ERDC/CHL and Max Agnew at USACE New Orleans District.

### Funding

The development of the USACE's CHS (https://chs.erdc.dren.mil), the CHS Probabilistic Framework (CHS-PF), the CHS Compound Coastal-Inland Framework (CHS-CF), the Joint Probability Method Aided by Metamodel Prediction (JPM-AMP), and related datasets have been funded in part by CHS multi-agency initiative (USACE's Civil Works R&D Programs, FEMA, U.S. Nuclear Regulatory Commission). This study was supported by the USACE's Engineer Research and Development Center, Coastal and Hydraulics Laboratory (ERDC-CHL) under contract award number ID W912HZ20C0050. All opinions expressed in this paper are the authors' and do not necessarily reflect the policies and views of the research sponsor (USACE) or any other organization.

**Competing Interests**

The authors have no relevant financial or non-financial interests to disclose.

**Author Contributions**

ZL: Conceptualization, Methodology, Software, Investigation, Formal analysis, Data Curation, Visualization, Writing—Original Draft, Writing—Review & Editing. MC: Conceptualization, Methodology, Visualization, Validation, Project administration, Writing—Review & Editing, Supervision. NNC: Data Curation, Writing—Review & Editing, Supervision, Funding Acquisition. LA: Data Curation, Writing—Review & Editing. MY: Data Curation, Writing—Review & Editing. MB: Conceptualization, Methodology, Software, Investigation, Formal analysis, Visualization, Supervision, Writing—Original Draft, Writing—Review & Editing.

## Appendix A Discretization information of BNs

**Table A-1. The discretization information of the Tier-1 BN nodes**

| NO. | Node | Discretized values |
|---|---|---|
| 1 | $\Delta P$ (hPa) | 8, 28, 48, 68, 88, 108, 128 |
| 2 | $V_\mathrm{f}$ (km/hr) | 0,11.67, 23.33, 35, 46.67, 58.33 |
| 3 | $R_{\max}$ (km) | 8, 40, 72, 104, 136, 168 |
| 4 | $Q$ (kcfs) | 0, 150, 300,… 1350 (equal interval) |
| 5 | $\hat{Z}$ (m) | -0.5, -0.23, 0.05, …, 10.23 (equal interval) |
| 6 | $Z$ (m) | -0.5, -0.23, 0.05, …, 10.23 (equal interval) |
| 7 | $\varepsilon_z$ | -∞,-3,-2,-1,0,1,2,3 |
| 8 | $Z_\mathrm{t}$ (m) | -0.5, -0.23, 0.05, …, 10.23 (equal interval) |
| 9 | $T$ | -∞,-3,-2,-1,0,1,2,3 |
| 10 | $X_0$ | See **Figure A-2** |
| 11 | $\Theta$ (deg) | -80, -60, -40, -20, 0, 20, 40, 60 |
| 12 | $S$ | Season 1, Season 2 |

Note: The discretized values presented in rows 1 to 10 provide the lower edge (value) of each bin.

**Table A-2. The Discretization information of the Tier-2 BN nodes**

| NO. | Node | Discretized values |
|---|---|---|
| 1 | $I$ | LI, MI, HI |

Note: Nodes not listed use the same discretization information as the Tier-1 BN.

**Table A-3. The Discretization information of the Tier-3 BN nodes**

| NO. | Node | Discretized values |
|---|---|---|
| 1 | $C$ ($°C$) | -0.05, 0.05, 0.15, 0.25, 0.35, 0.45 |
| 2 | $L$ (m) | -0.44, 0.44, 1.32, 2.2, 3.08, 3.96 |
| 3 | $P_a$ (mm) | 0, 41.67, 83.33, 125, 166.67, 208.33 |
| 4 | $Z_1$ (m) | -0.5, -0.13, 0.25,… 14.13 (equal interval) |
| 5 | $S_a$ | Season 1, Season 2 |

Note: The discretized values presented in rows 1 to 4 provide the lower edge (value) of each bin. Nodes not listed use the same discretization information as the Tier-1 BN and Tier-2 BN.

The discretization method of $X_0$ used in this study is the same as Liu et al. (2025). The location of representatives of $x_0$ and projected landfall locations are replicated in Figure A-1 and Figure A-2 for convenience.

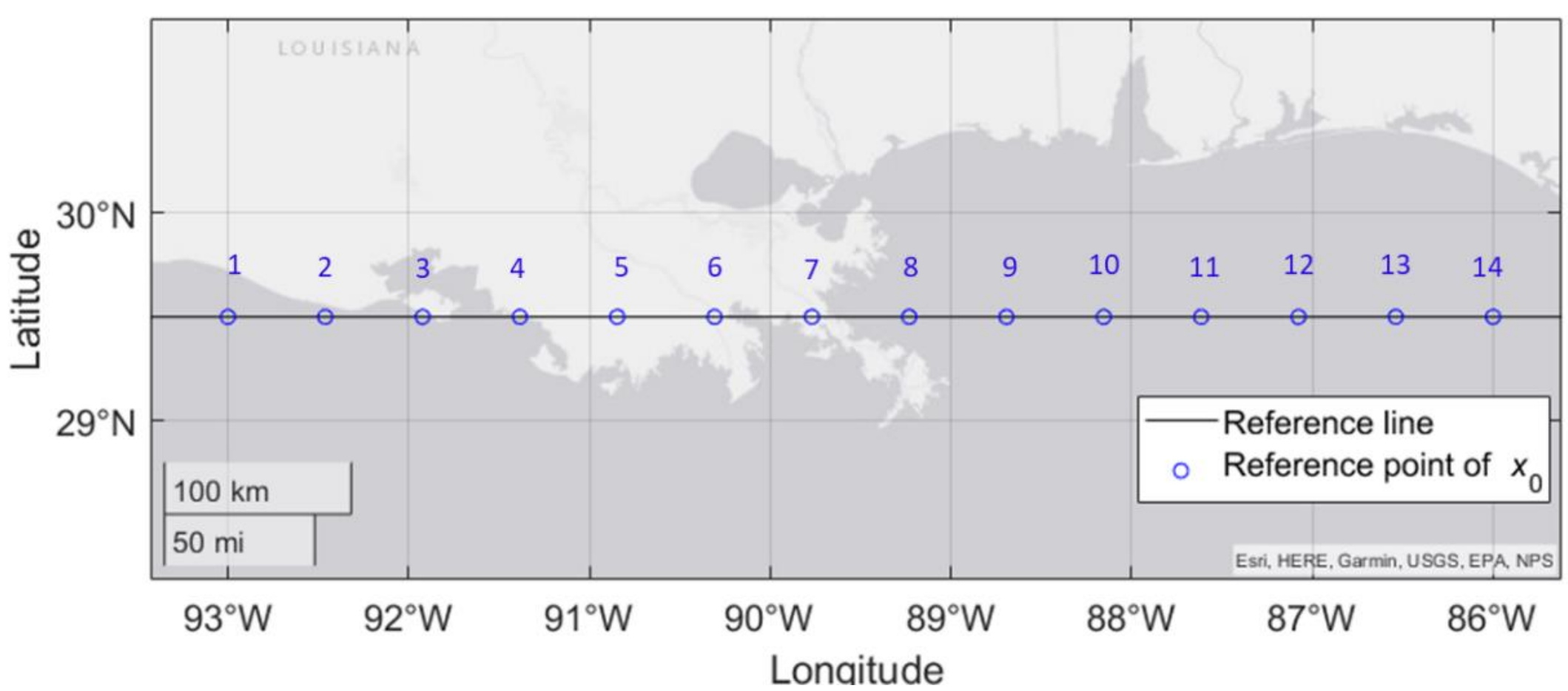

**Figure A-1: Representatives of $x_0$**

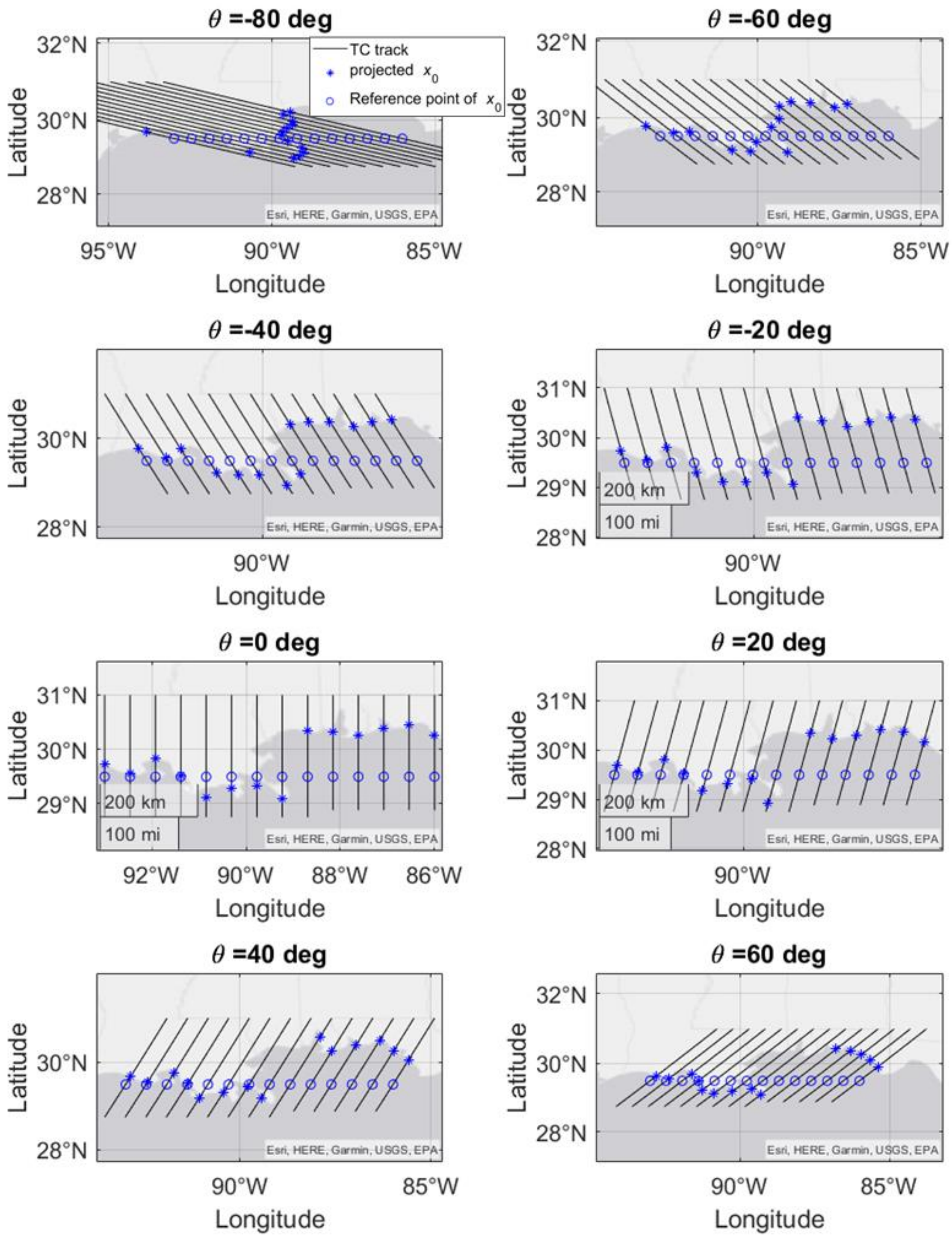

**Figure A-2: Projected landfall locations for specified values of $\Theta$**

# Appendix B Results for both sites

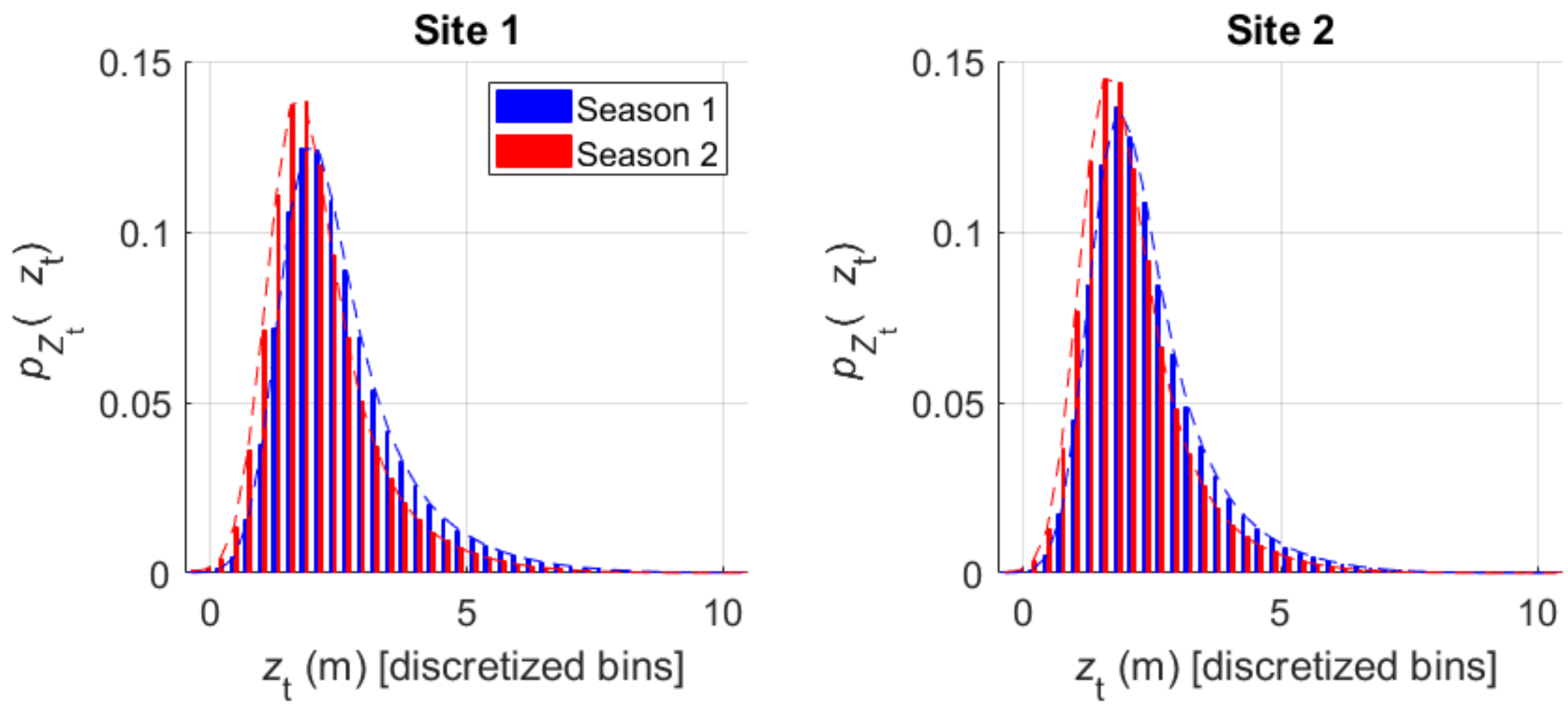

**Figure B-1: CCF WSL ($Z_t$) distribution generated by the Tier-1 BN**

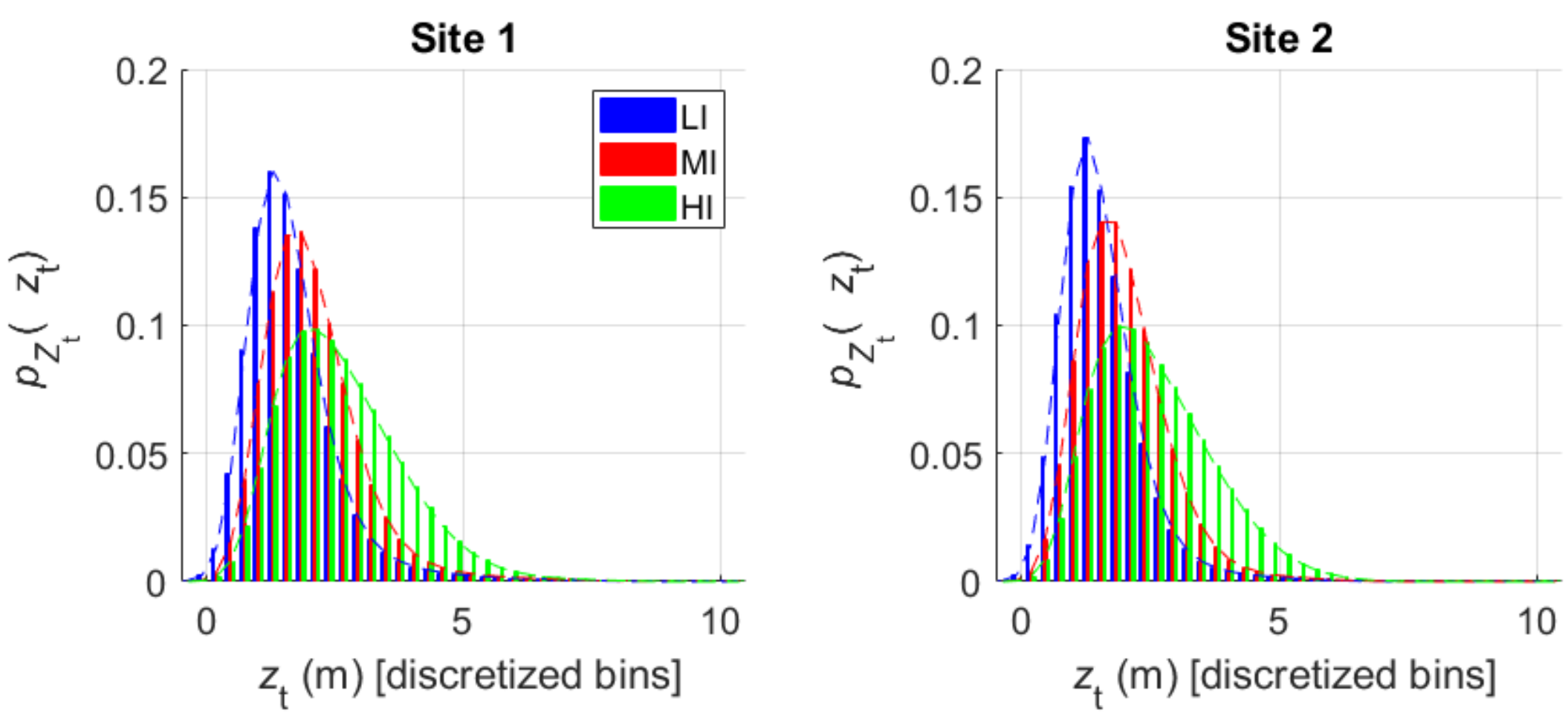

**Figure B-2: CCF WSL ($Z_t$) distribution generated by the Tier-2 BN**

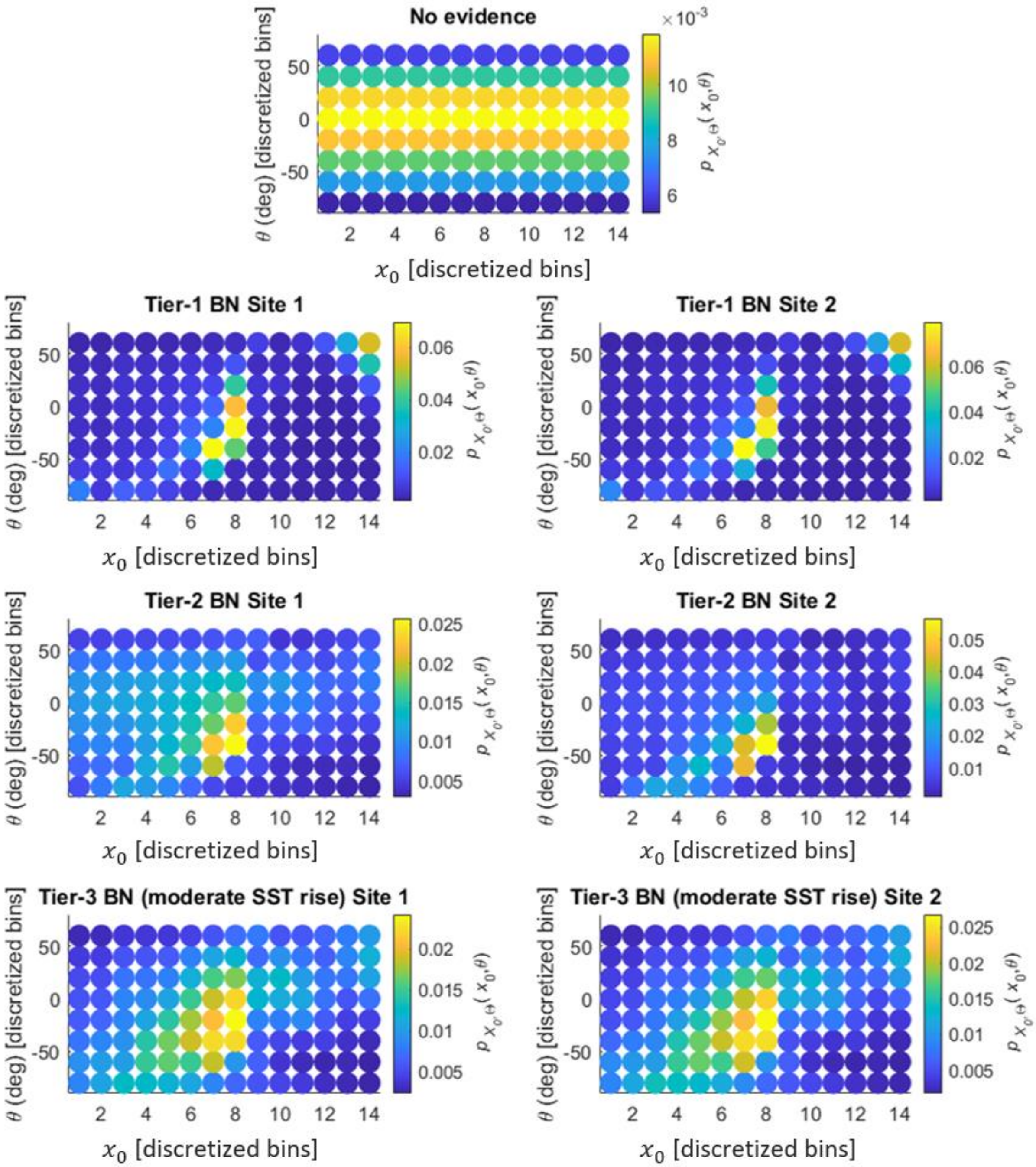

**Figure B-3: Pair-wise joint PM of $X_0$ and $\Theta$ under ECs (WSL > 5 m) at Site 1 (left column) and Site 2 (right column). Note that the range of color bars are different in each plot**

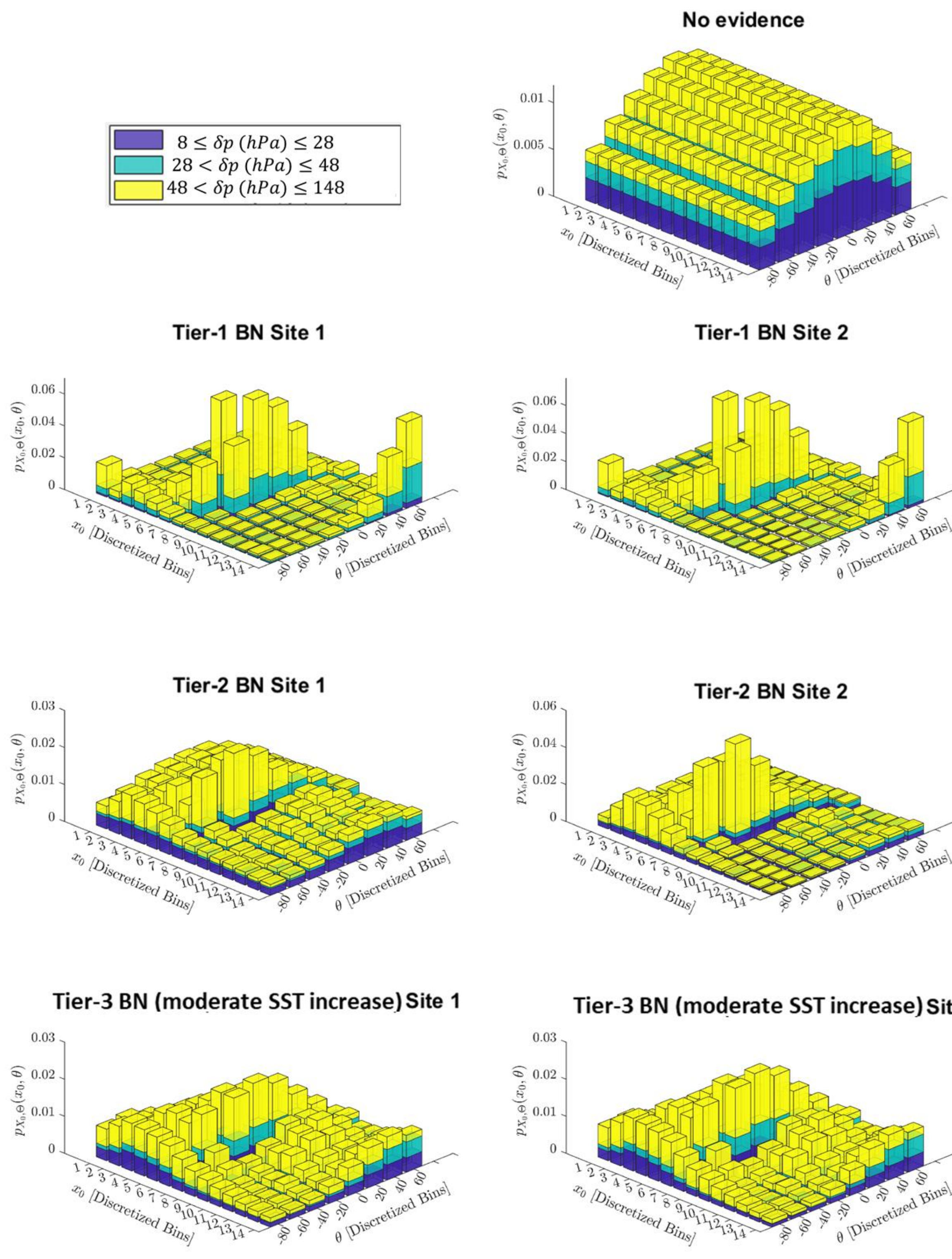

**Figure B-4: 3D stacked bar plot of TC track information with $\Delta P$ identified by colors**

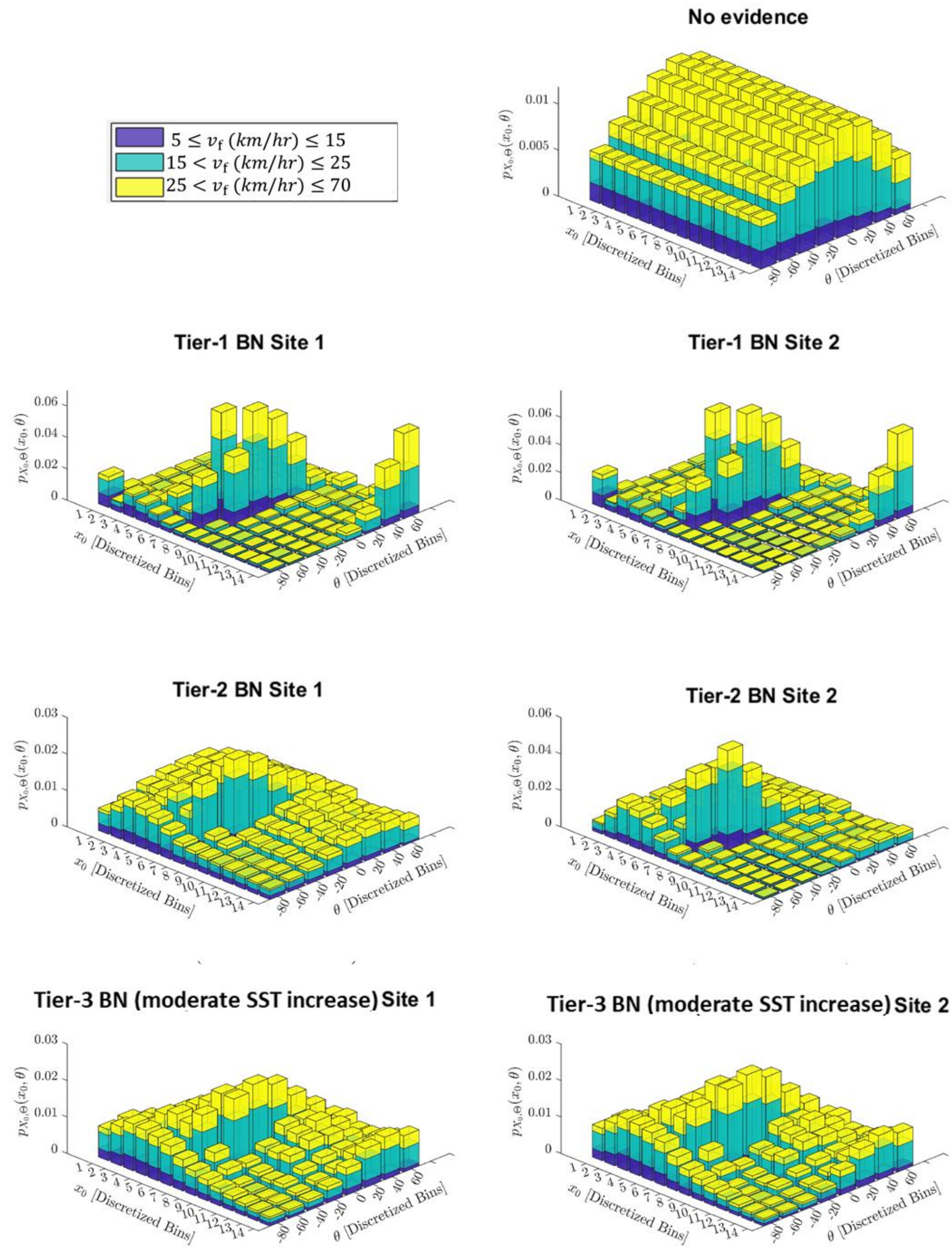

**Figure B-5: 3D stacked bar plot of TC track information with $V_{\mathrm{f}}$ identified by colors**

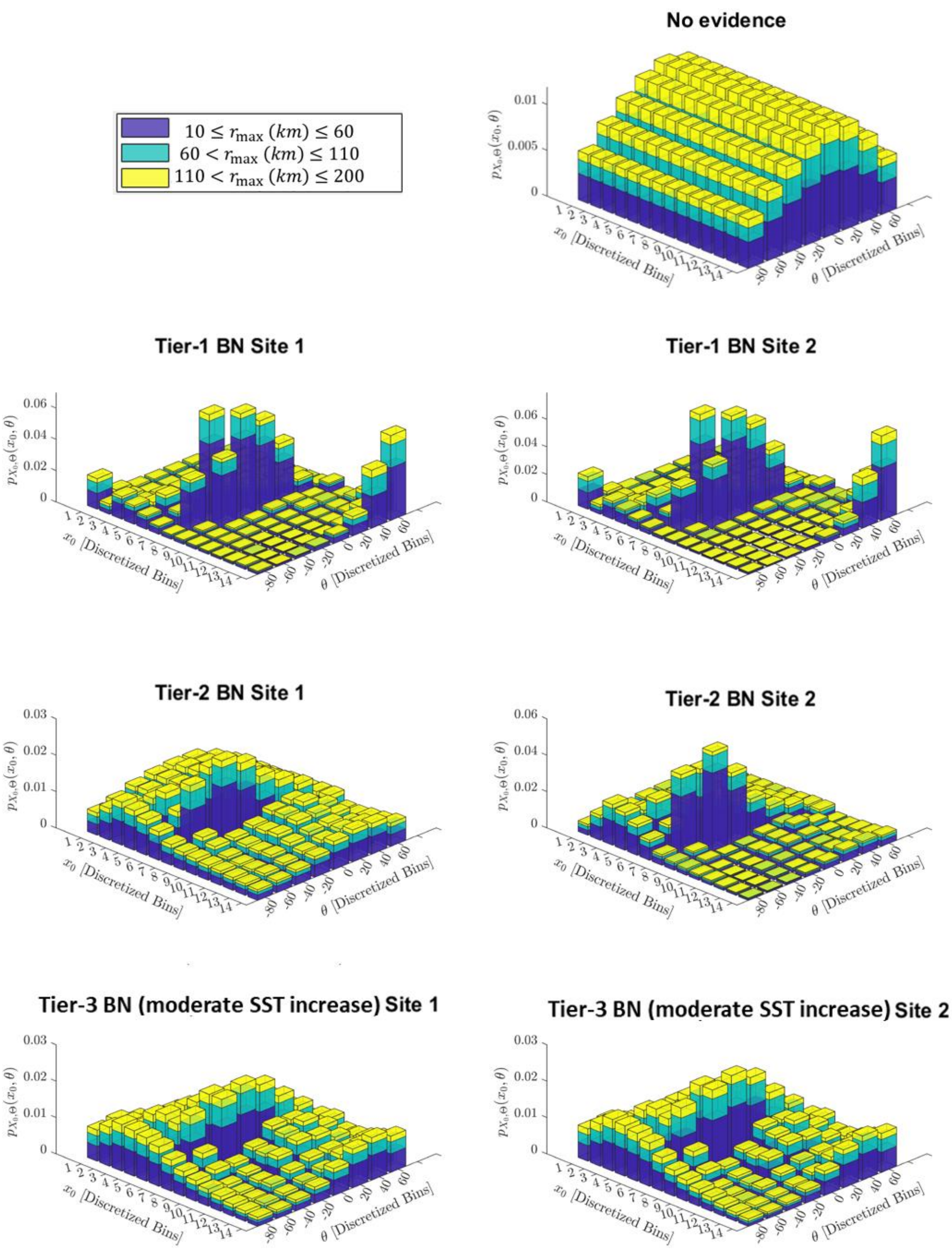

**Figure B-6: 3D stacked bar plot of TC track information with $R_{\max}$ identified by colors**